\documentclass[12pt]{article}
\pdfoutput=1

\usepackage{cite}
\usepackage{amsmath}
\usepackage{amssymb}
\usepackage{bm}

\usepackage{graphicx}

\title{}
\date{}
\def\para{\\ [-2mm]}

\def\para{\\ [-2mm]}
\def \be  {\begin{equation}}
\def \ee  {\end{equation}}
\def \ba  {\begin{eqnarray}}
\def \ea  {\end{eqnarray}}
\newcommand{\nn}{\nonumber}
\def\eqn#1{eq.~(\ref{#1})} 
\def\Eqn#1{Equation~(\ref{#1})}
\def\eqns#1#2{eqs.~(\ref{#1}) and~(\ref{#2})}
\def\Eqns#1#2{Eqs.~(\ref{#1}) and~(\ref{#2})}

\def\IZ{\relax\ifmmode\mathchoice
{\hbox{\cmss Z\kern-.4em Z}}{\hbox{\cmss Z\kern-.4em Z}}
{\lower.4pt\hbox{\cmsss Z\kern-.4em Z}}
{\lower1.2pt\hbox{\cmsss Z\kern-.4em Z}}\else{\cmss Z\kern-.4em Z}\fi}
\newcommand{\Z}{\mathsf{Z}\kern -5pt \mathsf{Z}}
\newcommand{\unit}{\mathsf{1}\kern -3pt \mathsf{l}}
\def\ie{{i.e.}}
\def\viz{{viz.}}

\def \Tr {\mathop{\rm Tr}\nolimits}

\def\half{  {\textstyle{1 \over 2}} }
\def\fr#1#2{ {\textstyle{#1 \over #2}}}

\def\eps{\epsilon}

\def\cA {  {\cal A} }

\def\< { \langle}
\def\> { \rangle}

\def\One {{ (1) }}
\def\Two{ {(2)} }
\def\Three{{(3)} }
\def\Four{ {(4)} }

\def\Ell{{(L)}}

\def\bP { {\bf P} }
\def\bQ { {\bf Q} }
\def\bc { {\bf c} }
\def\br { {\bf r} }
\def\bN { {\bf N} }
\def\bK { {\bf K} }
\def\blam { {\bm \lambda} }
\def\tp{\tilde{p}}
\def\tq{\tilde{q}}
\def\tP{\tilde{P}}
\def\tQ{\tilde{Q}}
\def\tf{\tilde{f}}

\def\sun{SU($N$)}
\def\son{SO($N$)}
\def\spn{Sp($N$)}
\def\T[#1]{ T_{[#1]} }
\def\oneN{ {1 \over N} }
\def\CC{   C }
\def\CCzero {\CC^{(0)}_{1234}  }
\def\CCone {\CC^{(1)}_{1234}  }
\def\CConeB{  \CC^{(1)}_{1342}  }
\def\CConeC{  \CC^{(1)}_{1423}   }
\def\CCtwoP  {\CC^{(2P)}_{1234}  }
\def\CCtwoNP  {\CC^{(2NP)}_{1234}  }

\def\xxi { x^i }
\def\xone { x^1 }
\def\xtwo { x^2 }
\def\AA{ A}
\def\BB{ B}
\def\sumsix{ \sum_{\la=1}^{6}   }
\def\sumla{ \sum_{\la}}
\def\ncon{n_{\rm null}}
\def\ncol{n_{\rm color}}
\def\ntr{n_{\rm trace}}
\def\aa{  m }
\def\la{  \lambda }
\def\Mila{ M_{i\la}^\Ell }
\def\Ala{ A_\la^\Ell }

\def\lam{\lambda}
\def\kap{\kappa}
\def\index{ c }

\begin{document}

\titlepage
\begin{flushright}
BOW-PH-172\\
\end{flushright}

\vspace{3mm}

\begin{center}

{\Large\bf\sf
All-loop group-theory constraints
\\ [2mm]
for four-point amplitudes of
\\ [2mm]
 \sun, \son, and \spn\  gauge theories
\\ [4mm]
}

\vskip 3cm

{\sc
Stephen G. Naculich
and 
Athis Osathapan
}

\vskip 0.5cm
{\it
Department of Physics and Astronomy\\
Bowdoin College\\
Brunswick, ME 04011 USA
}

\vspace{5mm}
{\tt
naculich@bowdoin.edu, aosathap@bowdoin.edu
}
\end{center}

\vskip 3cm

\begin{abstract}
In the decomposition of gauge-theory amplitudes into kinematic and color factors, the color factors (at a given loop order $L$) span a proper subspace of the extended trace space (which consists of single and multiple traces of generators of the gauge group, graded by powers of $N$).  Using an iterative process, we systematically construct the $L$-loop color space of four-point amplitudes of fields in the adjoint representation of SU($N$), SO($N$), or Sp($N$).  We define the null space as the orthogonal complement of the color space.  For SU($N$), we confirm the existence of four independent null vectors (for $L \ge 2$) and for SO($N$) and Sp($N$), we establish the existence of seventeen independent null vectors (for $L \ge 5$).  Each null vector corresponds to a group-theory constraint on the color-ordered amplitudes of the gauge theory. 
\end{abstract}

\vspace*{0.5cm}

\vfil\break

\section{Introduction}
\setcounter{equation}{0}

Gauge-theory scattering amplitudes at tree and loop level
may be represented in a gauge-invariant way 
in terms of color-ordered (or partial) 
amplitudes \cite{Mangano:1990by,Bern:1990ux}.
The color-ordered amplitudes for a particular process
are not independent but satisfy a number 
of constraints.
Some of these constraints are a consequence of 
color-kinematic duality \cite{Bern:2008qj,Bern:2010ue},
a property possessed by the amplitudes of a wide class of gauge theories,
whose most notable consequence is the  
gauge-gravity correspondence 
(see ref.~\cite{Bern:2019prr} for a comprehensive review).
Color-kinematic duality implies the existence of the
Bern-Carrasco-Johansson relations of tree-level amplitudes \cite{Bern:2008qj}
which were proven in 
refs.~\cite{BjerrumBohr:2009rd,Stieberger:2009hq,Feng:2010my,Chen:2011jxa}.
\para

There are, however, other constraints on color-ordered amplitudes that
are more basic because they follow directly from group theory,
such as the Kleiss-Kuijf relations 
among tree-level $n$-point amplitudes \cite{Kleiss:1988ne,DelDuca:1999rs},
the Bern-Kosower relations 
among one-loop \sun\ $n$-point amplitudes \cite{Bern:1994zx,DelDuca:1999rs},
and a two-loop relation that holds 
for four-point \sun\  color-ordered amplitudes \cite{Bern:2002tk}.
These group-theory relations for four-point \sun\ amplitudes were
generalized to all loop orders by one of the current 
authors using an iterative procedure \cite{Naculich:2011ep}.
This iterative technique was subsequently used by Edison and one
of the current authors to derive all-loop-order relations for
five-point \sun\ amplitudes \cite{Edison:2011ta}
and for six-point \sun\ amplitudes \cite{Edison:2012fn}. 
These results have been used in refs.~\cite{Geyer:2018xwu,Abreu:2018aqd,Chicherin:2018yne,Abreu:2019rpt,Badger:2019djh,Ahmed:2019qtg,Dunbar:2019fcq,Dalgleish:2020mof,Caron-Huot:2020vlo,DHoker:2020prr,Kosower:2022bfv,Agarwal:2023suw,DeLaurentis:2023nss}.
Other work on loop-level relations among $n$-point amplitudes
includes refs.~\cite{Kol:2014sla,Dunbar:2023ayw,DeLaurentis:2023izi}.
\para

The primary focus of this paper is to derive
all-loop-order group-theory constraints 
for four-point amplitudes of fields in the adjoint representation
of the classical groups \son\ and \spn, while also confirming the results
of ref.~\cite{Naculich:2011ep} for \sun.
While \sun\  is obviously most phenomenologically relevant
in the standard model context, 
\son\  and \spn\  could become relevant for theories
beyond the standard model,  e.g. grand unified theories.
In previous work, Huang \cite{Huang:2016iqf,Huang:2017ont}
generalized the iterative procedure of ref.~\cite{Naculich:2011ep}
to obtain group-theory constraints for four- and five-point amplitudes of 
\son\ and \spn\  up to four loops,
but did not uncover any patterns that could generalize to an arbitrary
number of loops.
\para

In this paper, we develop a 
refined version of the iterative approach 
that allows us obtain the all-loop structure
of the space of color factors
for all of the classical groups: \sun, \son, and \spn.
Not surprisingly, for \sun\  we rederive 
the {\it four} group-theory constraints for $L$-loop amplitudes (for $L \ge 2$)
obtained in ref.~\cite{Naculich:2011ep}.
For \son\ and \spn, we uncover a substantially more intricate structure
that implies the existence of {\it seventeen} group-theory constraints
for $L$-loop amplitudes (for $L \ge 5$).
(See tables \ref{table:sunnull} and \ref{table:sonnull} for
the number of constraints for all values of $L$.)
\para

Obtaining group-theory constraints for color-ordered amplitudes
boils down to a problem in linear algebra.
One begins with the amplitude (at some loop order $L$) 
expressed in a basis of color factors \cite{DelDuca:1999ha,DelDuca:1999rs}
\be
\cA^\Ell= \sum_i  a_i^\Ell \CC_i^\Ell
\label{colorbasis}
\ee
where $a_i^\Ell$ carries the momentum and polarization dependence 
of the amplitude,
and the color factors $\CC_i^\Ell$ are obtained by sewing together
group-theory factors from all the vertices of the contributing 
Feynman diagrams.
In a theory that contains only fields 
in the adjoint representation of the gauge group,
such as pure or supersymmetric Yang-Mills theory, 
each cubic vertex contributes a factor of the 
structure constants $\tf^{abc}$ of the gauge group $G$,
whereas each quartic vertex contributes a sum of products
of $\tf^{abc}$,
each of which are equivalent (from a purely color perspective)
to a pair of cubic vertices sewn along one leg.
Hence a complete set of color factors $\{ \CC_i^\Ell \}$ 
may be constructed from $L$-loop diagrams with cubic vertices only.
The color factors constructed from the set of {\it all} cubic diagrams 
are generally not independent but are related by Jacobi relations. 
We denote the number of independent color factors 
(\ie,  the dimension of the space of color factors)
as $\ncol$.
An independent basis of color factors
for tree-level and one-loop $n$-point amplitudes 
was described in refs.~\cite{DelDuca:1999ha,DelDuca:1999rs}.
One of our goals is obtain an independent basis of color factors
for four-point amplitudes at any loop order
for \sun, \son, and \spn.
\para

One may alternatively decompose the amplitude 
in a trace basis \cite{Mangano:1990by,Bern:1990ux}
\be
\cA^\Ell= \sumla  \Ala   t_\la^\Ell
\label{tracebasis}
\ee
whose coefficients are gauge-invariant color-ordered amplitudes $\Ala$
and the basis $\{  t_\la^\Ell \}$ consists of single and (at loop level) multiple 
traces of gauge group generators
$T^a$ in the defining  representation of the gauge group $G$.
The explicit form and dimensionality $\ntr$ of this 
(extended) trace basis depends on the gauge group.
For $G=$ \sun, one has $\ntr= 3L+3$ while for $G =$ \son\ or \spn, 
one has $\ntr=6L+3$.
The dimension of the trace basis is always larger than
that of the independent color basis ($\ntr > \ncol$) 
so there is redundancy among the color-ordered amplitudes,
expressed below as group-theory relations (\ref{rightrelations}). 
\para

The color (\ref{colorbasis}) and trace (\ref{tracebasis})
decompositions are related by writing the structure constants as 
\be
\tf^{abc} = \Tr( T^a, [T^b, T^c] )
\label{structure}
\ee
and then using group-dependent identities satisfied by the generators 
(see sec.~\ref{sec:tracebasis})
to express  each color factor $\CC_i^\Ell$ 
as a linear combination of trace factors
\be
\CC_i^\Ell =  \sumla \Mila t_\la^\Ell \,.
\label{trans}
\ee
Since $\ntr > \ncol$, 
the linear combinations given by \eqn{trans} 
span a proper subspace 
(which we will refer to as the color space)  
of the extended trace space. 
Consequently, the transformation matrix $\Mila$ possesses 
a set of independent null eigenvectors 
\be
\sumla  \Mila  r_{\la \aa}^\Ell = 0 \,, \qquad  \aa = 1, \cdots, \ncon 
\label{null}
\ee
whose number $\ncon$ is the difference between the dimensions of the trace
space and the color space.  
The null vectors, defined by 
$r_\aa^\Ell = \sumla r_{\la \aa}^\Ell t_\la^\Ell $,
are orthogonal to the color factors $\CC_i^\Ell$ 
with respect to the inner product 
\be 
(t_\la^\Ell, t_{\la'}^\Ell ) = \delta_{\la\la'} 
\label{innerproduct}
\ee 
and hence span the orthogonal complement of the color space; 
we refer to this as the {\it null space}.
\para

One combines \eqn{trans} with \eqns{colorbasis}{tracebasis}
to express the color-ordered amplitudes as
\be
\Ala = \sum_i  a_i^\Ell \Mila \,.
\label{colorordered}
\ee
Applying \eqn{null} to \eqn{colorordered}
implies the set of constraints
\be
\sumla  \Ala r_{\la \aa}^\Ell = 0 \,, \qquad \aa = 1, \cdots, \ncon
\label{rightrelations}
\ee
which we refer to as group-theory relations.
Hence, specifying the null space is equivalent to specifying the complete
set of group-theory relations satisfied by the color-ordered amplitudes.
\para

\begin{table}[t]
\begin{center}
\begin{tabular}{|l|*{7}{r}|c|}
\hline
number of loops &  0 & 1 & 2 & 3 & 4 & 5 & 6 &  $L\ge 2$ \\
\hline
$\ncol$ & 2 & 3 & 5 & 8 & 11 & 14  &  17 &  $3L-1$ \\
$\ntr$  & 3 & 6 & 9 & 12 & 15 & 18 & 21 &  $ 3L+3 $\\
$\ncon$ & 1 & 3 & 4 & 4 & 4 & 4 & 4 & 4 \\
\hline
\end{tabular}
\end{center}
\caption{
Dimensions of color, trace, and null spaces for \sun\ amplitudes}
\label{table:sunnull} 
\vskip .2 cm
\end{table}

\begin{table}[t]
\begin{center}
\begin{tabular}{|l|*{7}{r}|c|}
\hline
number of loops       & 0 & 1 & 2  & 3  & 4  & 5  & 6  &  $L\ge 5$ \\
\hline
$\ncol  $   & 2 & 3 & 5  & 8  & 11 & 16 & 22 & $6L-14$\\
$\ntr   $   & 3 & 9 & 15 & 21 & 27 & 33 & 39 & $6L+3$ \\
$\ncon$ &  1 & 6 & 10 & 13 & 16 & 17 & 17 & 17 \\
\hline
\end{tabular}
\end{center}
\caption{
Dimensions of color, trace, and null spaces for \son\ and \spn\ amplitudes}
\label{table:sonnull}
\end{table}

The iterative approach taken in this paper involves
attaching a rung across any pair of external legs of an
arbitrary $L$-loop color factor.   
We make the assumption that doing this to all diagrams
spanning the space of $L$-loop color factors 
generates the space of $(L+1)$-loop color factors,
an assumption borne out in practice.
Starting with the tree-level color space,
we explicitly construct a set of color factors
spanning the color space at each loop order 
for \sun, \son, and \spn.
In tables \ref{table:sunnull} and \ref{table:sonnull},
we list the dimensions of these color spaces,
together with the dimensions of trace and null spaces,
where $\ncon =\ntr-\ncol$.
The dimensions in table \ref{table:sunnull}
confirm the results of ref.~\cite{Naculich:2011ep} for
four-point \sun\ amplitudes at all loop orders.
The dimensions in table \ref{table:sonnull}
are in agreement with ref.~\cite{Huang:2016iqf}
for four-point \son\ and \spn\ amplitudes for $0 \le L \le 4$.
Ref.~\cite{Huang:2016iqf} did not go beyond four loops.
\para

Since the complete set of color factors at a given loop order is 
invariant under permutation of the external legs, 
the color space forms a representation of $S_4$,
which can be decomposed into irreducible representations
of one and two dimensions, denoted in this paper by $u$ and $x$ respectively.
The trace and null spaces also decompose into $u$- and $x$-type 
irreducible representations.
For \sun, there are generically (for $L \ge 2$) four null vectors,
two of $u$-type and two of $x$-type,  
for which we determine the explicit forms. 
For \son\ and \spn, there are generically (for $L \ge 5$) seventeen
null vectors, seven of $u$-type and ten of $x$-type.
We determine (for arbitrary $L$) 
the explicit forms of the ten $x$-type null vectors in this paper,
leaving the seven $u$-type null vectors to future work.
\para 

This paper is structured as follows.
In sec.~\ref{sec:tracebasis} we review the color and trace spaces
for $L$-loop four-point amplitudes through two loops,
decomposing them into irreducible representations of $S_4$.
In sec.~\ref{sec:iterative} we review and refine the iterative procedure
for  generating the $(L+1)$-loop color space from the $L$-loop color space. 
In sec.~\ref{sec:suncolor} we employ this refined iterative procedure
to generate the $L$-loop color space for \sun. 
In sec.~\ref{sec:sunnull} after defining an inner product
on the trace space, we determine the $L$-loop null space for \sun,
the orthogonal complement of the $L$-loop color space with respect
to this inner product.
In sec.~\ref{sec:soncolor} we generate 
the $L$-loop color space for \son, 
and in sec.~\ref{sec:sonnull} 
we obtain the complete set of $x$-type null vectors for \son. 
Sec.~\ref{sec:spncolor} briefly explains how the results
from \spn\ are related to those of \son. 
Sec.~\ref{sec:concl} concludes the paper,
and some technical details are 
relegated to two appendices.
\para

\section{Trace and color spaces}
\label{sec:tracebasis}
\setcounter{equation}{0}

In this section, 
we describe in some detail 
the trace and color spaces associated
with color factors for \sun, \son, and \spn\  four-point amplitudes
through two loops.
This will set the stage for the subsequent discussion of all-loop color
factors in the remainder of the paper.
First, we describe the decomposition 
of $L$-loop color factors into the trace basis for each group.
The span of these color factors gives the $L$-loop color space.
We then break these color spaces into irreducible representations 
of $S_4$, the permutation group of the external legs of the amplitude,
which allows for the most efficient representation of these spaces.

\subsection{Trace basis decomposition of low-loop color factors}

\begin{figure}[t]
\begin{center}
\includegraphics[width=4.0cm]{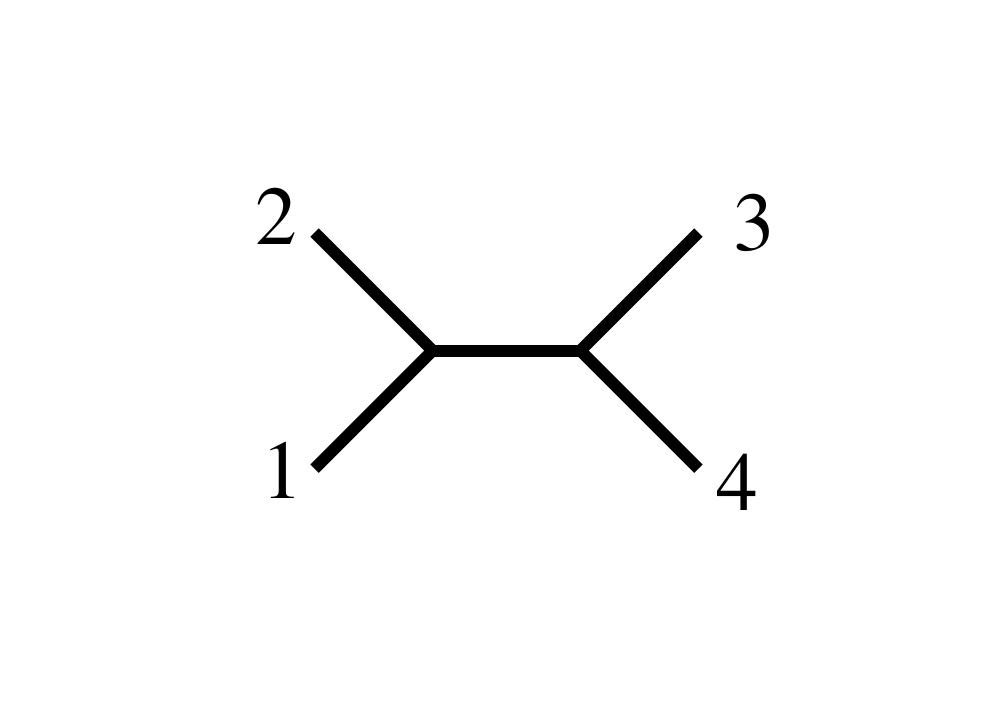}
\includegraphics[width=4.0cm]{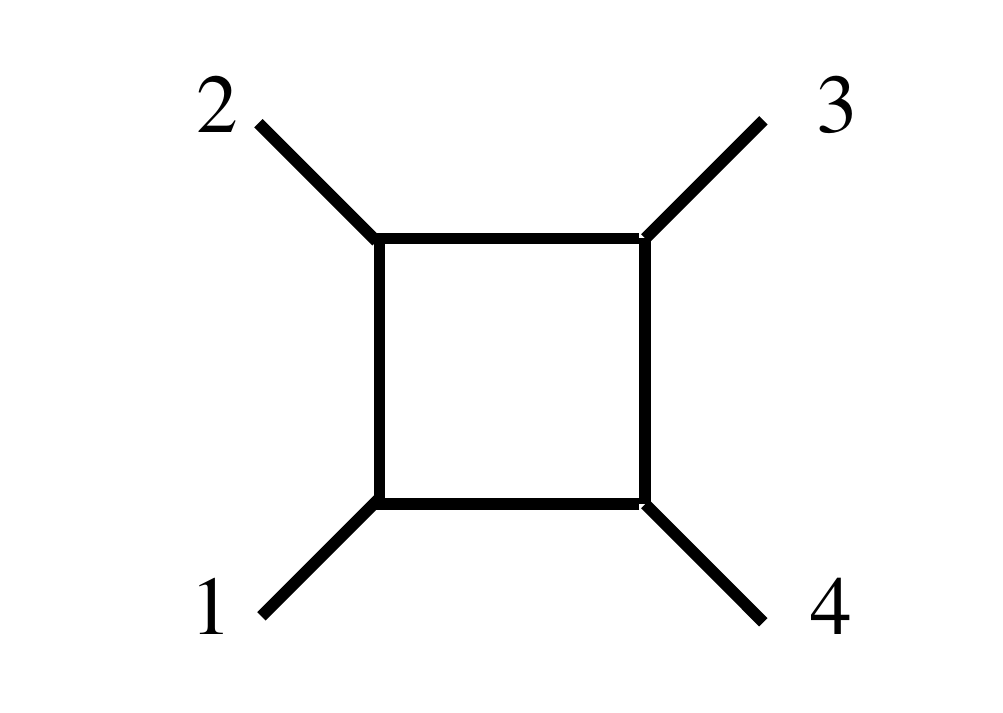}
\includegraphics[width=4.0cm]{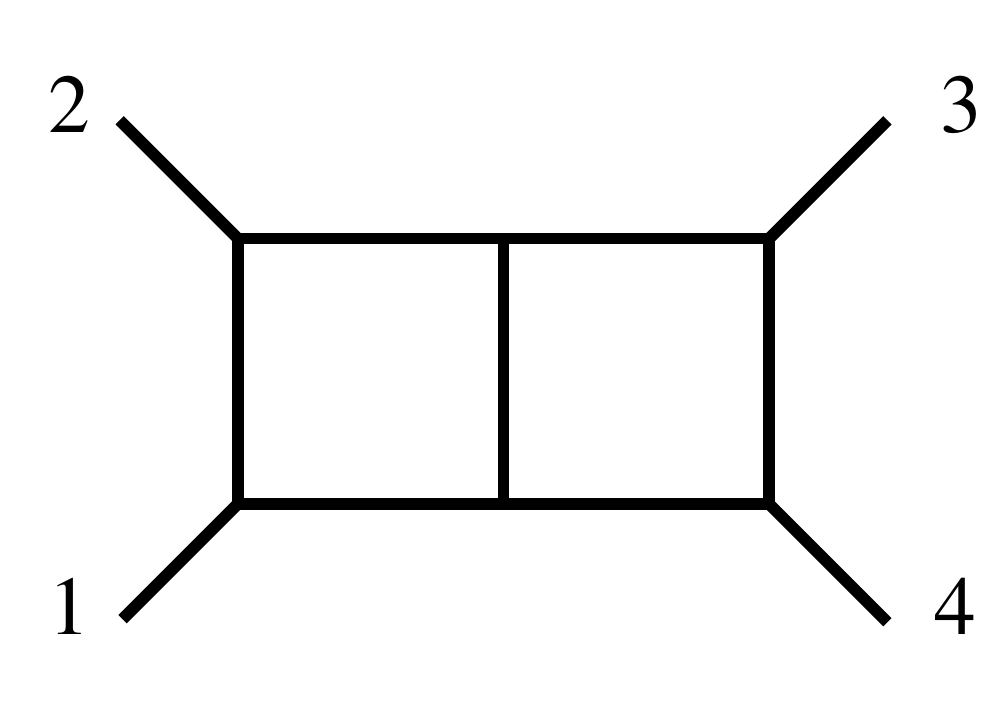}
\hskip5mm
\includegraphics[width=4.0cm]{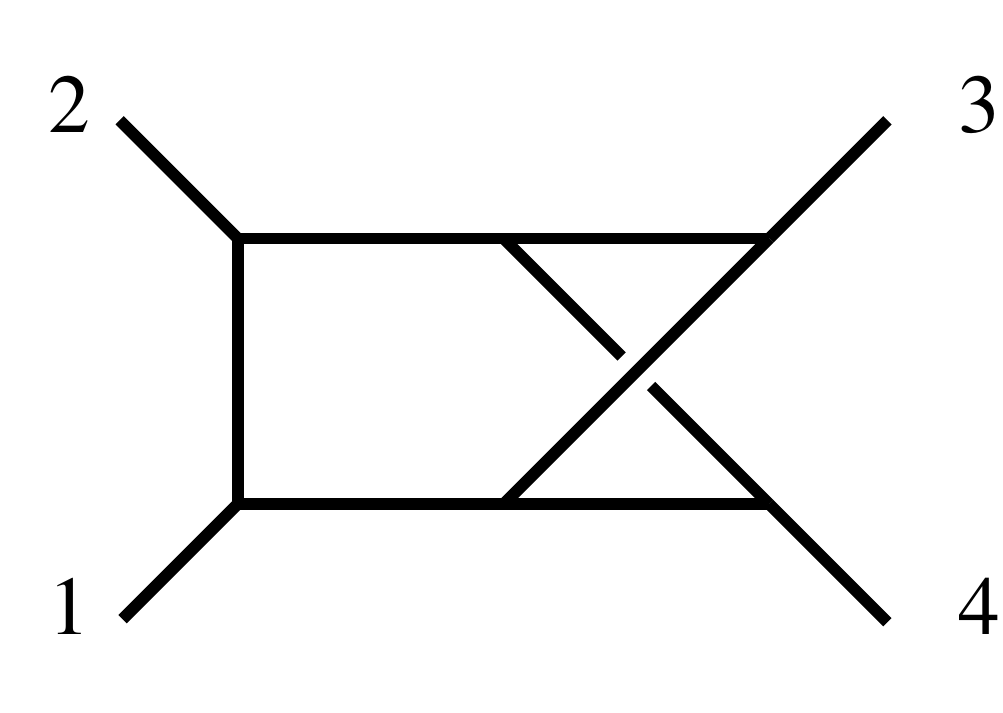}
\caption{Tree-level, one-loop, and two-loop four-point color factors} 
\label{fig:one}
\end{center}
\end{figure}

Color factors for amplitudes of fields in the adjoint representation 
are constructed by contracting structure constants $\tf^{abc}$
of the associated group.   
For example, for the four-point diagrams shown in fig.~\ref{fig:one},
the $s$-channel tree-level color factor is given by 
\begin{align}
\CCzero = \tf^{a_1 a_2 e} \tf^{a_3 a_4 e} \,,
\label{schannel}
\end{align}
the one-loop box color factor is
\begin{align}
\CCone = \tf^{e a_1 b} \tf^{b a_2 c} \tf^{c a_3 d} \tf^{d a_4 e}  \,,
\label{oneloopbox}
\end{align}
and the two-loop planar and nonplanar color factors are
\begin{align}
\CCtwoP
&=\tf^{e a_1 b} \tf^{b a_2 c} \tf^{cgd} \tf^{dfe}\tf^{g a_3 h}\tf^{h a_4 f}\,,
\label{twoloopP}\\
\CCtwoNP 
&=\tf^{e a_1 b} \tf^{b a_2 c} \tf^{cgd} \tf^{hfe}\tf^{g a_3 h}\tf^{d a_4 f} \,.
\label{twoloopNP}
\end{align}

The decomposition of these color factors into the trace basis is accomplished
by rewriting the structure constants in terms of 
generators $T^a$ in the defining representation using
\begin{align}
\tf^{abc} = \Tr( T^a,[T^b, T^c] ) \,,  \qquad\qquad
[T^a, T^b] = \tf^{abc} T^c \,, \qquad\qquad
\Tr(T^a T^b) = \delta^{ab} \,.
\label{fabc}
\end{align}

By repeatedly using trace identities specific to each gauge group 
(as described below),
one can reduce all color factors to a linear combination
of traces and products of traces of generators.  
In particular, four-point color factors may be written in terms of a
six-dimensional basis $\T[\lam]$ of single and double traces of generators
\begin{align}
\CC = \sum_{\lam=1}^6 \CC_{[\lam]}  ~\T[\lam] \,.
\label{sixdimdecomp}
\end{align}

For SU($N$), we will use the following 
basis\footnote{We are following the convention of ref.~\cite{Edison:2012fn} 
rather than that of ref.~\cite{Naculich:2011ep}. 
By including factors of two in the double-trace terms,
this basis generalizes more naturally 
to the trace basis for higher-point amplitudes \cite{Edison:2012fn}.\label{fn1}}
\begin{align}
\T[1] &=  
\Tr(T^{a_1} T^{a_2} T^{a_3} T^{a_4}) + \Tr(T^{a_1} T^{a_4} T^{a_3} T^{a_2}),
\qquad
\T[4] =  2 \Tr(T^{a_1} T^{a_3}) \Tr(T^{a_2} T^{a_4}) , 
\nn\\
\T[2] &= 
 \Tr(T^{a_1} T^{a_3} T^{a_4} T^{a_2})+ \Tr(T^{a_1} T^{a_2} T^{a_4} T^{a_3}), 
\qquad
\T[5] =  2 \Tr(T^{a_1} T^{a_4}) \Tr(T^{a_2} T^{a_3})  ,
\nn \\
\T[3] &=  
\Tr(T^{a_1} T^{a_4} T^{a_2} T^{a_3}) + \Tr(T^{a_1} T^{a_3} T^{a_2} T^{a_4}),
\qquad
\T[6] =  2 \Tr(T^{a_1} T^{a_2}) \Tr(T^{a_3} T^{a_4})\,.  
\label{suntracebasis} 
\end{align}

For \son\ and \spn, 
the trace of a product $\BB$ of generators is equal (up to a possible sign)
to the trace of the generators written in reverse order $\BB^R$ 
\begin{align}
\Tr(\BB^R) &= (-1)^{n_\BB} \Tr(\BB) & 
\hbox{for \son~ and \spn}
\label{reversal}
\end{align}
where $n_\BB$ denotes the number of factors in $\BB$,
using \eqns{sonR}{spnR} in appendix \ref{sec:appA}.
This implies that for \son\ and \spn, the trace basis (\ref{suntracebasis})
simplifies 
to\footnote{We retain the factors of 2
for consistency with \eqn{suntracebasis},
but they may easily be removed.}
\begin{align}
\T[1] &=  
2 \Tr(T^{a_1} T^{a_2} T^{a_3} T^{a_4}) \,,
\qquad
\qquad
\qquad
\qquad
\T[4] = 2 \Tr(T^{a_1} T^{a_3}) \Tr(T^{a_2} T^{a_4}) \,, 
\nn\\
\T[2] &= 
2 \Tr(T^{a_1} T^{a_3} T^{a_4} T^{a_2}) \,,
\qquad
\qquad
\qquad
\qquad
\T[5] = 2 \Tr(T^{a_1} T^{a_4}) \Tr(T^{a_2} T^{a_3})  \,,
\nn \\
\T[3] &=  
2  \Tr(T^{a_1} T^{a_4} T^{a_2} T^{a_3}) \,,
\qquad
\qquad
\qquad
\qquad
\T[6] = 2 \Tr(T^{a_1} T^{a_2}) \Tr(T^{a_3} T^{a_4})  \,.
\label{sonspntracebasis} 
\end{align}
Note that $\Tr(T^a)=0$ for all the groups considered,
so that there are no other terms in the four-point trace basis.
\para

We now describe the process of decomposing the color factors 
shown in fig.~\ref{fig:one}
into the trace basis (\ref{suntracebasis}) for \sun\ 
and (\ref{sonspntracebasis}) for \son\ and \spn.
For all groups $G$, 
the tree-level color factor (\ref{schannel}) reduces to 
\begin{align}
\CCzero 
=  \Tr( T^{a_1}, [T^{a_2}, T^e]) \tf^{a_3 a_4 e} 
 =  \Tr( T^{a_1}, [T^{a_2}, [T^{a_3}, T^{a_4} ]] )  
 = \T[1]- \T[2]  
\label{treelevel}
\end{align}
where we have used \eqn{fabc}.
Similarly, the one-loop color factor (\ref{oneloopbox}) becomes
\begin{align}
\CCone = 
\Tr( T^{e}, [T^{a_1}, T^{b}] ) \tf^{b a_2 c} \tf^{c a_3 d} \tf^{d a_4 e}
= \Tr( T^{e}, [T^{a_1}, [T^{a_2}, [T^{a_3}, [T^{a_4}, T^e]]]] )  
\label{oneloopexpand}
\end{align}

where we are left with a contraction over $T^e$.
The remainder of the calculation depends on the group $G$.
For \sun, one uses the identities (\ref{sunrln})
valid for generators in the defining representation
\begin{align}
\Tr(\AA T^a) \Tr(\BB T^a)
& = \Tr (\AA\BB) - \oneN \Tr(\AA) \Tr(\BB) \,,
\nn\\
\Tr(\AA T^a \BB T^a)
&= \Tr (\AA) \Tr(\BB) - \oneN \Tr(\AA \BB) 
\end{align}

where $\AA$ and $\BB$ are arbitrary products of generators. 
Then \eqn{oneloopexpand} yields
\begin{align}
\CCone
= N \T[1] + \T[4]+\T[5]+\T[6] 
\qquad \hbox{for  \sun } \,.
\label{oneloopsun}
\end{align}

For \son\ and \spn, one uses instead the identities 
(\ref{sonrln}) and (\ref{spnrln})
\begin{align}
\Tr(\AA T^a) \Tr(\BB T^a)
& = \frac{1}{2} \Big[ \Tr (\AA\BB) - (-1)^{n_\BB} \Tr(\AA \BB^R)\Big] \,,
\nn\\[2mm]
\Tr(\AA T^a \BB T^a)
&= \frac{1}{2} \Big[ \Tr (\AA) \Tr(\BB) \mp (-1)^{n_\BB} \Tr(\AA \BB^R) \Big]
\end{align}

in which case \eqn{oneloopexpand} reduces to 
\begin{align}
\CCone =
  \half (N \mp 4) \T[1] 
  \mp   \Big( \T[2] + \T[3]\Big) 
+ \half \Big( \T[4]+\T[5]+\T[6] \Big) 
\qquad \hbox{for \son\  and  \spn} \,.
\label{oneloopson}
\end{align}

The two-loop color factors (\ref{twoloopP}) and (\ref{twoloopNP})
may be similarly reduced to the six-dimensional trace basis in this way.
\para

It is convenient to represent a color factor
in the trace basis (\ref{sixdimdecomp})
as a six-dimensional row vector 
\begin{align}
\CC 
=( \CC_{[1]},~ \CC_{[2]},~ \CC_{[3]};~ \CC_{[4]},~ \CC_{[5]},~ \CC_{[6]})  \,.
\label{rowvector}
\end{align}

Thus for \sun, the tree-level, one-loop,
and two-loop color factors are represented as 
\begin{align}
\CCzero &=  ( 1,~ -1,~  0;~  0,~  0,~  0) \,,\nn\\
\CCone &= (N,~  0,~  0;~  1,~  1,~  1) \,,\nn\\
\CCtwoP  &=  (N^2 + 2,~  2,~  -4;~  0,~  0,~  3N) \,,\nn\\
\CCtwoNP &= (2,~  2,~ -4;~  -N,~  -N,~  2N)  \,.
\label{lowloopsun}
\end{align}

For \son, the tree-level, one-loop,
and two-loop color factors are represented as 
\begin{align}
\CCzero &=  ( 1,~ -1,~  0;~  0,~  0,~  0) \,,\nn\\
\CCone &= \half (N- 4,~  - 2,~  - 2;~  1,~ 1,~ 1) \,,\nn\\
\CCtwoP &=\fr{1}{4} (N^2 - 7N + 16,~  - 3N + 12,~  -12;~ 2,~ 2,~ 3N - 10)  \,,\nn\\
\CCtwoNP &= \fr{1}{4} 
(- N+8,~  - N+8,~   2N-16;~  -N + 4,~  -N + 4,~  2N - 8)
\,.
\label{lowloopson}
\end{align}

For \spn, the tree-level, one-loop, and two-loop color factors are
represented as
\begin{align}
\CCzero &=  ( 1,~ -1,~  0;~  0,~  0,~  0) \,, \nn\\
\CCone &= \half (N+ 4,~   2,~   2;~  1,~ 1,~ 1)  \,,\nn\\
\CCtwoP &=\fr{1}{4} (N^2 + 7N + 16,~   3N + 12,~  -12;~  - 2,~  - 2,~  3N + 10)   \,,\nn\\
\CCtwoNP &= \fr{1}{4} 
( N+8,~   N+8,~  - 2N-16;~  -N - 4,~  -N - 4,~  2N + 8)
\,.
\label{lowloopspn}
\end{align}

We will use these results to decompose the color spaces 
into irreducible representations of $S_4$.

\subsection{Trace space and color space}

As one can see from the low-loop examples 
(\ref{lowloopsun})-(\ref{lowloopspn})
in the previous subsection,
in an $L$-loop color factor of the form (\ref{rowvector}),
the first three terms $\CC_{[1]}$, $\CC_{[2]}$, and $\CC_{[3]}$
are polynomials in $N$ of maximal degree $L$ 
and the second three terms $ \CC_{[4]}$, $ \CC_{[5]}$, and $ \CC_{[6]} $
are polynomials in $N$ of maximal degree $L-1$.
Furthermore, for \sun, 
$\CC_{[1]}$, $\CC_{[2]}$, and $\CC_{[3]}$
are polynomials of even/odd degree depending on whether $L$ is even/odd,
and vice versa for $ \CC_{[4]}$, $ \CC_{[5]}$, and $ \CC_{[6]} $.
\para

Color factors can be regarded as belonging to a
vector space $V^\Ell$, 
which we call the $L$-loop trace space,
consisting of all such polynomials.
Of course, an $L^{\rm th}$ degree polynomial may be regarded as an element 
of an $(L+1)$-dimensional vector space, whose components are given 
by the coefficients of the polynomial\footnote{If the polynomial is even
or odd, the vector space has dimension $\lceil {L+1 \over 2} \rceil$.}.
Thus the dimension of the $L$-loop trace space is
\begin{align}
\dim  V^\Ell =
\begin{cases}
3L+3 & \hbox{ for \sun}, \\
6L+3 & \hbox{ for \son\ and \spn} \,.
\end{cases}
\end{align}
In ref.~\cite{Naculich:2011ep}, 
we defined an explicit basis for the trace space of \sun,
called the extended trace basis $t^\Ell_\lam$,
whose elements were of the form $N^n \T[\lam]$.
Similarly, an extended trace basis for \son\ and \spn\ was defined in
ref.~\cite{Huang:2016iqf}.
In this paper, it is more convenient to express color factors 
in polynomial form.
\para

The set of all $L$-loop color factors, 
formed from all possible cubic diagrams,
spans a proper subspace of $V^\Ell$.
We call this subspace the $L$-loop color space.
In the color space we must include 
all permutations of external legs
of the color factors.
For example, the tree-level color space includes not only
the $s$-channel diagram shown in fig.~\ref{fig:one},
but the $t$- and $u$-channel diagrams
obtained by permutations of the external legs.
Given the trace decomposition (\ref{rowvector})
of a particular cubic diagram, 
the trace decompositions of the same color factor with
permutations of the external legs  are given by 
\begin{align}
\CC_{1234}  &=   ( \CC_{[1]},~ \CC_{[2]},~ \CC_{[3]}; ~\CC_{[4]},~ \CC_{[5]},~ \CC_{[6]} ) \,,
\nn\\[1mm]
\CC_{1243}  &=   ( \CC_{[2]},~ \CC_{[1]},~ \CC_{[3]}; ~\CC_{[5]},~ \CC_{[4]},~ \CC_{[6]} ) \,,
\nn\\[1mm]
\CC_{1342}  &=   ( \CC_{[3]},~ \CC_{[1]},~ \CC_{[2]}; ~\CC_{[6]},~ \CC_{[4]},~ \CC_{[5]} ) \,,
\nn\\[1mm]
\CC_{1324}  &=   ( \CC_{[3]},~ \CC_{[2]},~ \CC_{[1]}; ~\CC_{[6]},~ \CC_{[5]},~ \CC_{[4]} ) \,,
\nn\\[1mm]
\CC_{1423}  &=   ( \CC_{[2]},~ \CC_{[3]},~ \CC_{[1]}; ~\CC_{[5]},~ \CC_{[6]},~ \CC_{[4]} ) \,,
\nn\\[1mm]
\CC_{1432}  &=   ( \CC_{[1]},~ \CC_{[3]},~ \CC_{[2]}; ~\CC_{[4]},~ \CC_{[6]},~ \CC_{[5]} ) 
\label{otherperms}
\end{align}

as may easily be seen by examining \eqns{suntracebasis}{sonspntracebasis}.
\subsection{Irreducible subspaces}
\label{subsec:irred}

Because the set of all possible $L$-loop cubic diagrams 
necessarily includes all permutations of the external legs,
the $L$-loop color space forms a (reducible) representation of $S_4$, 
the group of permutations of the external legs.
This representation can be reduced to a set of 
irreducible one- and two-dimensional representations in the form
of Kronecker products
\begin{align}
[P, Q] \otimes u, \qquad\qquad  [P, Q] \otimes \xxi, \quad (i=1,2)
\end{align}

where
\begin{align}
u = ( 1,1,1), \qquad\qquad
\xone = (1,-1,0), \qquad\qquad 
\xtwo = (0,1,-1), 
\label{xudef}
\end{align}

and $P$ and $Q$ are polynomials in $N$
of maximal degree $L$  and $L-1$ respectively. 
That is,
\begin{align}
[P, Q] \otimes u  
&\equiv (P, P, P; Q, Q, Q)  \,,
\nn\\
[P,Q] \otimes \xone
&\equiv (P, -P, 0; Q, -Q, 0)\,,  \nn\\
[P,Q] \otimes \xtwo
&\equiv (0, P, -P; 0, Q, -Q)  \,.
\end{align}
The decomposition of the single- and double-trace bases 
into irreducible representations of $S_n$ was described
in detail in ref.~\cite{Edison:2012fn}.
\para

An arbitrary color factor (\ref{rowvector}) may be decomposed into
irreducible representations of $S_4$ as follows: 
\begin{align}
\CC =  &\fr{1}{3} \Big(
~[\CC_{[1]} + \CC_{[2]} + \CC_{[3]},  \CC_{[4]} + \CC_{[5]} + \CC_{[6]}] 
\otimes u 
\nn\\
&+ 
[ 2\CC_{[1]} - \CC_{[2]} - \CC_{[3]}, 2 \CC_{[4]} - \CC_{[5]} - \CC_{[6]}] 
\otimes \xone
\nn\\[1mm]
&+
[ \CC_{[1]} + \CC_{[2]} -2  \CC_{[3]},  \CC_{[4]} + \CC_{[5]} -2 \CC_{[6]}] 
\otimes \xtwo
\Big)
\end{align}
as is easily verified using \eqn{xudef}.
For example, 
the tree-level color factor (\ref{schannel}) and its permutations
are given by
\begin{align}
\CCzero &=  ( 1,-1, 0; 0, 0, 0) = [1,0] \otimes \xone \,,\nn\\
\CC^{(0)}_{1342}  
        &=  ( 0, 1,-1; 0, 0, 0) = [1,0] \otimes \xtwo \,,\nn\\
\CC^{(0)}_{1423}  
        &=  (-1, 0, 1; 0, 0, 0) = [1,0] \otimes (- \xone - \xtwo) \,.
\end{align}
These three color factors thus span a 2-dimensional representation
$[1,0] \otimes \xxi $ of $S_4$.
(They are not independent due to the Jacobi identity
$\CCzero + \CC^{(0)}_{1342}  + \CC^{(0)}_{1423}  =0$.)
The one-loop \sun\ color factor 
$\CCone =  ( N, 0,  0; 1, 1, 1)$
decomposes into 
\begin{align}
\CCone &= \fr{1}{3}  [N,3] \otimes u 
        + \fr{2}{3} [N, 0] \otimes \xone 
        + \fr{1}{3} [N, 0] \otimes \xtwo  \,.
\end{align}
This color factor and its permutations
\begin{align}
\CConeB &=  ( 0, N, 0; 1, 1, 1) \,, \nn\\
\CConeC &=  ( 0, 0, N; 1, 1, 1) 
\end{align}
span a 3-dimensional representation of $S_4$
which reduces to 
a 1-dimensional representation 
$[N,3]\otimes u$  and a 2-dimensional representation
$[N,0] \otimes \xxi$.
The two-loop planar and nonplanar 
\sun\ color factors (\ref{lowloopsun}) decompose into 
\begin{align}
\CCtwoP  &= \fr{1}{3} [N^2, 3N] \otimes u 
        + \fr{1}{3} [2N^2+6, -3N] \otimes \xone 
        + \fr{1}{3} [N^2 + 12, -6N] \otimes \xtwo\,,  \nn\\[1mm]
\CCtwoNP &= [2, -N] \otimes \xone + [4, -2N] \otimes \xtwo 
\end{align}
so that the two-loop color space consists of
the 1-dimensional representation $[N^2, 3N] \otimes u$
and two 2-dimensional representations
$[N^2,0] \otimes \xxi$ and $[2,-N] \otimes \xxi$. 
\para

Summarizing our results, we see that the low-loop \sun\ color spaces
are spanned by 
\begin{align}
\hbox{Tree-level: }
& [1,0] \otimes \xxi\,,  \nn\\[2mm]
\hbox{One-loop: }
& [N,3] \otimes u \,, \nn\\
&  [N,0] \otimes \xxi  \,,
\nn\\[2mm]
\hbox{Two-loop: }
& [N^2, 3N] \otimes u \,, \nn\\
& [N^2,0] \otimes \xxi \,, \nn\\
& [2,-N] \otimes \xxi  \,.
\label{suntreeonetwo}
\end{align}

The same procedure employed for the \son\ color factor spaces yields
\begin{align}
\hbox{Tree-level: }
& [1,0] \otimes \xxi \,, \nn\\[2mm]
\hbox{One-loop: }
& [N-8,3] \otimes u \,, \nn\\
&  [N-2,0] \otimes \xxi \,,  
\nn\\[2mm]
\hbox{Two-loop: }
& [(N-2)(N-8) , 3(N-2)] \otimes u \,, \nn\\
& [(N-2)^2,0] \otimes \xxi \,, \nn\\
& [N-8,N-4] \otimes \xxi 
\label{sontreeonetwo}
\end{align}

and for the \spn\ color spaces
\begin{align}
\hbox{Tree-level: }
& [1,0] \otimes \xxi \,, \nn\\[2mm]
\hbox{One-loop: }
& [N+8,3] \otimes u \,, \nn\\
&  [N+2,0] \otimes \xxi  \,,
\nn\\[2mm]
\hbox{Two-loop: }
& [(N+2)(N+8) , 3(N+2)] \otimes u \,, \nn\\
& [(N+2)^2,0] \otimes \xxi \,, \nn\\
& [N+8,N+4] \otimes \xxi  \,.
\label{spntreeonetwo}
\end{align}

These results will be useful in generating the color space
for an arbitrary loop amplitude.
\para

We observe that the dimensions of the color spaces are 
2, 3, and 5 for $L=0$, 1, and 2, respectively for all three groups, 
as reflected in tables \ref{table:sunnull} and \ref{table:sonnull}.
We will see below that this equality between the groups breaks 
down for $L \ge 5$.
\para

\section{Iterative procedure} 
\label{sec:iterative} 
\setcounter{equation}{0}

In this section, we review the iterative procedure
introduced by one of the current authors 
in ref.~\cite{Naculich:2011ep} to generate a complete
set of color factors at all loop orders. 
We then present a refined version of the iterative approach
that takes into account the decomposition of 
color spaces into irreducible representations of $S_4$.
\para

The iterative approach involves attaching a rung between any two external legs
of an $L$-loop color factor to generate an $(L+1)$-loop color factor.
By considering all possible attachments of rungs, 
one generates the space of color factors at $(L+1)$ loops. 
The orthogonal complement of the color space in the trace space
defines the null space, \ie, the space of null eigenvectors 
of the transformation matrix (\ref{null}).
Each null eigenvector then corresponds 
to a group-theory constraint on the color-ordered amplitudes. 
\para

Given an $L$-loop color factor $\CC^{a_1 a_2 a_3 a_4}$,
attaching a rung between external legs 1 and 2 yields an 
$(L+1)$-loop color factor given by 
\begin{align}
\CC^{a_1 a_2 a_3 a_4} \longrightarrow
\tf^{a_1 b_1 c} \tf^{c b_2 a_2} \CC^{b_1 b_2 a_3 a_4} 
\end{align}
with similar expressions for the color factors obtained by
attachments of rungs between other legs.
To determine the effect of attaching rungs to 
an arbitrary color factor,
we define an iterative matrix $G (e_{12}, e_{13}, e_{14})$
by attaching rungs between different pairs of legs 
of the trace basis $T_{[\lam]}^{a_1 a_2 a_3 a_4}$ 
and decomposing the result in the trace basis 
\begin{align}
  e_{12} \tf^{a_1 b_1 c} \tf^{c b_2 a_2} \T[\lam]^{b_1 b_2 a_3 a_4} 
+ e_{13} \tf^{a_1 b_1 c} \tf^{c b_3 a_3} \T[\lam]^{b_1 a_2 b_3 a_4} 
&+e_{14} \tf^{a_1 b_1 c} \tf^{c b_4 a_4} \T[\lam]^{b_1 a_2 a_3 b_4} 
\\
&=  \sum_{\kap} G_{\lam \kap} (e_{12}, e_{13},e_{14})
\T[\kap]^{a_1 a_2 a_3 a_4} \,.
\nn
\end{align}

Thus the coefficient of $e_{12}$ gives the result of attaching a rung between
legs 1 and 2, etc.  
(We need not consider the effect of attaching rungs between
legs 2 and 3, etc., as they are redundant.) 
The $6 \times 6$ matrix $G_{\lam\kap}$
can be written in block diagonal form, with the $N$
dependence made explicit:
\begin{align}
G(e_{12}, e_{13}, e_{14}) &= 
\begin{pmatrix}
N A +E & B \\ C & N D + F
\end{pmatrix}
\label{defG}
\end{align}
where $A$ through $F$ are $3 \times 3$ matrices that depend on 
$e_{1j}$.
For \sun, 
one finds\footnote{These matrices differ slightly from those
in ref.~\cite{Naculich:2011ep} because of the factors 
of two multiplying the double-trace basis elements.  See footnote \ref{fn1}.}
\begin{align}
A &=
\begin{pmatrix}
e_{12}+e_{14} & 0 & 0 \\
0 & e_{12}+e_{13} & 0 \\
0 & 0 & e_{13}+e_{14} \\
\end{pmatrix} \,,
&B&=
\begin{pmatrix}
0 &  e_{14}- e_{13} &  e_{12}- e_{13} \\
 e_{13}- e_{14} & 0 &  e_{12}- e_{14} \\
 e_{13}- e_{12} &  e_{14}- e_{12} & 0 \\
\end{pmatrix} \,,
\nn\\
C &=
2 \begin{pmatrix}
0 &  e_{12}-e_{14} & e_{14}-e_{12} \\
e_{12}-e_{13} & 0 & e_{13}-e_{12} \\
e_{14}-e_{13} & e_{13}-e_{14} & 0 \\
\end{pmatrix}\,,
&D &=
2 \begin{pmatrix}
 e_{13} & 0 & 0 \\
0 &  e_{14} & 0 \\
0 & 0 &  e_{12} \\
\end{pmatrix}\,,
\label{sunG}
\end{align}

with $E$ and $F$ vanishing.
For \son\ (upper sign)  and \spn\ (lower sign), 
one finds
\begin{align}
A &=
{1 \over 2} 
\begin{pmatrix}
e_{12}+e_{14} & 0 & 0 \\
0 & e_{12}+e_{13} & 0 \\
0 & 0 & e_{13}+e_{14} \\
\end{pmatrix} 
\,,
~~~B=
{1 \over 2} \begin{pmatrix}
0 &  e_{14}- e_{13} &  e_{12}- e_{13} \\
 e_{13}- e_{14} & 0 &  e_{12}- e_{14} \\
 e_{13}- e_{12} &  e_{14}- e_{12} & 0 \\
\end{pmatrix} \,,
\nn\\
C &=
2 \begin{pmatrix}
0 &  e_{12}-e_{14} & e_{14}-e_{12} \\
e_{12}-e_{13} & 0 & e_{13}-e_{12} \\
e_{14}-e_{13} & e_{13}-e_{14} & 0 \\
\end{pmatrix}\,,
~~~~D =
\begin{pmatrix}
e_{13} & 0 & 0 \\
0 & e_{14} & 0 \\
0 & 0 & e_{12} \\
\end{pmatrix}\,,
\nn
\\
E &=
\pm {1 \over 2} \begin{pmatrix}
 2e_{13}- 3e_{12}-3e_{14} & e_{13}-e_{12} & e_{13}-e_{14} \\
e_{14}-e_{12} & 2e_{14}- 3e_{12}-3e_{13} & e_{14}-e_{13} \\
e_{12}-e_{14} & e_{12}-e_{13} & 2e_{12}- 3e_{13}-3e_{14} \\
\end{pmatrix}\,,
\nn\\
F & =
\mp 2 \begin{pmatrix}
 e_{13} & 0 & 0 \\
 0 &  e_{14} & 0 \\
 0 & 0 &  e_{12} \\
\end{pmatrix}\,.
\label{sonspnG}
\end{align}

The effect of attaching rungs to an arbitrary color factor 
$\CC = \sum_{\lam} \CC_{[\lam]}  \T[\lam]$
results in
\begin{align}
\CC_{[\lam]} \longrightarrow
\sum_{\kap} \CC_{[\kap]} ~G_{\kap\lam} (e_{12}, e_{13}, e_{14}) 
\end{align}

that is, one multiplies the row vector $\CC$ by the matrix $G$.
To give some simple examples,
attaching a rung between legs 1 and 4 of the 
tree-level $s$-channel diagram (\ref{schannel})
yields the one-loop box diagram (\ref{oneloopbox})
so that
\begin{align}
\CCzero G(0,0,1) &= \CCone 
\end{align}
while attaching a rung between legs 1 and 2 of the one-loop box
diagram yields the two-loop planar diagram (\ref{twoloopP})
so that 
\begin{align}
\CCone  G(1,0,0) &= \CCtwoP \,.
\end{align}
These may be confirmed 
using eqs.~(\ref{suntreeonetwo})-(\ref{spntreeonetwo})
and (\ref{defG})-(\ref{sonspnG}).

\subsection{Iterative matrices for irreducible representations of $S_4$}

We explained in sec.~\ref{subsec:irred} how color factors 
may be written in terms of irreducible representations
of $S_4$:
\begin{align}
[P, Q] \otimes u, \qquad\qquad  [P, Q] \otimes \xxi  \qquad (i=1,2) \,.
\end{align}
In general $G(e_{12}, e_{13}, e_{14})$ will act on these color factors
to produce linear combinations of $u$ and $\xxi$ types,
but we may define $G$ matrices for certain choices 
of the parameters $e_{12}$, $e_{13}$, and $e_{14}$ 
that produce pure $u$ and $\xxi$ types.
One may then write the action of $G$
in terms of four $2 \times 2$ matrices $g_1$, $g_{ux}$, $g_{xu}$, and $g_{xx}$
which act on the two-dimensional row vector $[P,Q]$.
This gives a refined approach to generate the color space for any $L$
in terms of $u$- and $\xxi$-type irreducible representations.
\para

First we choose $e_{12}=e_{13}=e_{14}= \half e$
which makes $G(e_{12}, e_{13}, e_{14})$ 
proportional to the unit matrix, 
mapping $u$-type color factors to $u$-type, 
and $\xxi$-type to $\xxi$-type:
\begin{align}
\left([P,Q] \otimes u\right) G(\half e, \half e, \half e) 
&= [P,Q] g_1 \otimes u   \,, \nn\\
\left([P,Q] \otimes \xxi\right) G(\half e, \half e, \half e) 
&= [P,Q] g_1 \otimes \xxi  \,.
\end{align}

One may verify that 
\begin{align}
g_1 &= e \begin{pmatrix} N & 0 \cr 0 & N \end{pmatrix}
\quad\hbox{for \sun},
&g_1 &= \half e \begin{pmatrix} N \mp 2 & 0 \cr 0 & N \mp 2 \end{pmatrix} 
\quad \hbox{for} \begin{cases} $\son$ \cr $\spn$ \end{cases}
\,.
\end{align}

Another choice of $e_{1j}$ takes $u$-type color factors to $\xxi$-type
color factors:
\begin{align}
\left([P,Q] \otimes u\right) G(0,-e,e) 
&= [P,Q] g_{ux} \otimes \xone \,, \nn\\
\left([P,Q] \otimes u\right) G(e,0,-e) 
&= [P,Q] g_{ux} \otimes \xtwo\,.  
\end{align}

In this case, one finds 
\begin{align}
g_{ux} &=e \begin{pmatrix} N & -3 \cr 6 & -2N \end{pmatrix} 
\quad\hbox{for \sun},
& g_{ux} &= \half e 
\begin{pmatrix} N \mp 5  & -3  \cr 12  & -2 N \pm 4 \end{pmatrix}
\quad \hbox{for} \begin{cases} $\son$ \cr $\spn$ \end{cases}
\,.
\end{align}

Yet another choice of $e_{1j}$ 
takes $\xxi$-type color factors to $\xxi$-type: 
\begin{align}
\left([P,Q] \otimes \xone\right) G(e,0,0)  
&= [P,Q] g_{xx} \otimes \xone \,, \nn\\
\left([P,Q] \otimes \xtwo\right) G(0,e,0)  
&= [P,Q] g_{xx} \otimes \xtwo \,. 
\end{align}

One then obtains
\begin{align}
g_{xx} &=e \begin{pmatrix} N & 0 \cr -2 & 0 \end{pmatrix} 
\quad\hbox{for \sun},
&g_{xx} &= \half e \begin{pmatrix} N\mp 2  & 0 \cr -4 & 0 \end{pmatrix} 
\quad \hbox{for} \begin{cases} $\son$ \cr $\spn$ \end{cases}
\,.
\end{align}

Finally one must act on the two $\xxi$-type color factors with 
different choices of $e_{1j}$ to obtain a $u$-type color factor
\begin{align}
\left([P,Q] \otimes \xone\right) G(0,-e,e) 
+\left([P,Q] \otimes \xtwo\right) G(e,-e,0) 
&=
[P,Q] g_{xu} \otimes u \,.
\end{align}

One then obtains
\begin{align}
g_{xu} &=e \begin{pmatrix} N & 3 \cr 0 & -2N \end{pmatrix} 
\quad\hbox{for \sun},
&g_{xu}&= \half e \begin{pmatrix} N \mp 8  & 3 \cr 0 & -2 N\pm 4 \end{pmatrix} 
\quad \hbox{for} \begin{cases} $\son$ \cr $\spn$ \end{cases}
\,.
\end{align}

The iterative matrices 
$g_1$, $g_{ux}$, $g_{xu}$, and $g_{xx}$
will be used to generate the 
$L$-loop color spaces for \sun\ in sec.~\ref{sec:suncolor}
and for \son\ in sec.~\ref{sec:soncolor}.
In sec.~\ref{sec:spncolor}, we will show that the
$L$-loop color spaces for \spn\ are obtained from those for \son\ 
by some simple sign changes.
\para

\section{$L$-loop \sun\ color space}
\label{sec:suncolor}
\setcounter{equation}{0}

The goal of this section is to explicitly construct the 
space of $L$-loop color factors for \sun.
As already discussed in sec.~\ref{subsec:irred},
an $L$-loop color factor may be expressed 
in terms of one- and two-dimensional irreducible representations 
of $S_4$ as 
\begin{align}
[P, Q] \otimes u, \qquad\qquad  [P, Q] \otimes \xxi \qquad (i=1,2)
\end{align}
where $P$ and $Q$ are polynomials in $N$
of maximal degree $L$  and $L-1$ respectively. 
For \sun\ color factors, the polynomials $P$ are of even/odd degree
depending on whether $L$ is even/odd,
and vice versa for $Q$. 
Thus, $L$-loop \sun\ color factors inhabit a vector space $V^\Ell$ 
of dimension $3L+3$ (the trace space). 
The polynomials $P$ and $Q$
corresponding to color factors, however,
are not completely arbitrary but satisfy certain constraints. 
Consequently, the set of all $L$-loop color factors 
spans a proper subspace (the color space) of $V^\Ell$.
\para

In this section, 
we iteratively construct an explicit basis 
for the $L$-loop \sun\ color space,
beginning with the 
single tree-level irreducible representation $[1,0]\otimes \xxi$
and acting repeatedly with the iterative matrices for \sun\ obtained
in sec.~\ref{sec:iterative}
\begin{align}
g_1 &= \begin{pmatrix} N & 0 \cr 0 & N \end{pmatrix}, \qquad
g_{xx} = \begin{pmatrix} N & 0 \cr -2 & 0 \end{pmatrix}, \qquad
g_{xu} = \begin{pmatrix} N & 3 \cr 0 & -2N \end{pmatrix}, \qquad
g_{ux} = \begin{pmatrix} N & -3 \cr 6 & -2N \end{pmatrix}
\label{gsun}
\end{align}

where we have chosen to set $e=1$.
These  $2 \times 2$ matrices act on the $[P,Q]$ part of the color factor,
while the subscripts indicate their action on the $x$ or $u$ part of the color factor.
Specifically:

\begin{enumerate}
\item
$g_{xx}$ takes an $x$-type color factor to an $x$-type color factor,

\item
$g_{xu}$ takes an $x$-type color factor to a $u$-type color factor,

\item
$g_{ux}$ takes a $u$-type color factor to an $x$-type color factor,

\item
$g_1$ takes $x$ to $x$ and $u$ to $u$. 
\end{enumerate} 

We will show that these matrices generate a basis 
consisting of polynomials multiplied by
one of four specific (linearly independent) types: 
\begin{align}
x_a &\equiv [1,0]\otimes \xxi \,,  \nn \\[1mm]
x_b &\equiv [2,-N]\otimes \xxi \,, \nn \\[1mm]
u_a &\equiv [N,3]\otimes u \,,  \nn \\[1mm]
u_b &\equiv [N,N^2+3]\otimes u \,. 
\label{suntypes}
\end{align}
Hints of these types have already appeared in \eqn{suntreeonetwo}.
Our first step is to ascertain how each of the operators (\ref{gsun})
act on the types of color factors (\ref{suntypes}).
First, the operator $g_1$ just rescales each type by $N$
\begin{align}
x_a  g_1 &= N x_a \,,\nn\\
x_b  g_1 &= N x_b \,,\nn\\
u_a  g_1 &= N u_a \,,\nn\\
u_b  g_1 &= N u_b  \,.
\label{sung1}
\end{align}

Second, the operator $g_{xx}$ acts on the $x$-type color factors as
\begin{align}
x_a  g_{xx} &= N x_a \,,\nn\\
x_b  g_{xx} &= 4N x_a  \,.
\end{align}
Since the action of $g_{xx}$ on $x_a$ is identical to the action of $g_1$
(and therefore redundant), 
we will restrict our attention to its action on $x_b$, defining 
$g_{ba} = \fr{1}{4} g_{xx}$ with
\begin{align}
x_b g_{ba} &= N  x_a  \,.
\label{sungba}
\end{align}

Third, the operator $g_{xu}$ takes an $x_a$-type color factor 
to a $u_a$-type color factor,
and  an $x_b$-type color factor to a $u_b$-type color factor:
\begin{align}
x_a g_{xu} &= u_a  \,,\nn \\
x_b g_{xu} &= 2 u_b  \,.
\label{sungxu}
\end{align}

Finally, the operator $g_{ux}$ acts on $u$-type color factors to give 
linear combinations of $x_a$ and $x_b$ types: 
\begin{align}
u_a g_{ux} &= 
N^2 \, x_a   + 9 x_b
\,,\nn \\
u_b g_{ux} &= 
3N^2 x_a  + (2N^2+9)x_b
\,.
\label{sungux}
\end{align}

With these in hand, we now generate the \sun\ color space 
through three-loop order. 
We begin with 
the single tree-level irreducible representation
\begin{align}
\hbox{Tree level:}\qquad     x_a \,.
\label{suntreelevel}
\end{align}

Acting on $x_a$ with $g_1$ using \eqn{sung1} 
and with $g_{xu}$ using \eqn{sungxu}, 
we obtain the three-dimensional space spanned by
two irreducible representations
\begin{align}
\hbox{One loop:}\qquad  N  x_a, \quad  u_a \,.
\label{sunoneloop}
\end{align}

We then act on each of these one-loop color factors 
with $g_1$ to obtain $N^2 x_a $ and $N u_a$. 
The action of $g_{xu}$ on $N x_a$ is redundant,
but we can act on the $u_a$-type color factor 
with $g_{ux}$ to obtain $[N^2+18,-9N] \otimes \xxi$,
which is a linear combination of $x_a$ and $x_b$ types,
as shown in \eqn{sungux}.
Since we already have $N^2 x_a$ in the color space, we subtract 
it and divide by $9$ to obtain  $x_b$.
Thus  the two-loop color space is five-dimensional, spanned by
three irreducible representations
\begin{align}
\hbox{Two loops:}\qquad  
N^2 x_a \,, 
\quad N u_a \,,
\quad x_b \,.
\label{suntwoloop}
\end{align}

It is reassuring that \eqns{sunoneloop}{suntwoloop} agree 
with the results we obtained earlier in \eqn{suntreeonetwo}.
The three-loop color factors are then obtained by acting 
on each of the two-loop color factors with $g_1$. 
The action of $g_{ux}$ on $N u_a$ is redundant.
We can also act on $x_b$
with $g_{ba}$ using \eqn{sungba}  and with $g_{xu}$ using \eqn{sungxu}.
The three-loop color space is thus eight-dimensional, spanned by 
five irreducible representations
\begin{align}
\hbox{Three loops:}\qquad  
N^3 x_a, 
\quad N x_a,
\quad N^2 u_a,
\quad N  x_b, 
\quad u_b \,.
\end{align}

We now make some general observations
that allow us to determine the complete span of color factors
at arbitrary loop order $L$.
\para

{\bf (Observation 1) All $L$-loop  $u$-type color factors are generated 
by the action of $g_{xu}$ on the complete set of $x$-type 
color factors at $(L-1)$ loops using \eqn{sungxu}.}
The only possible exception would be through $g_1$ acting on an
$(L-1)$-loop $u$-type color factor.
But since (by hypothesis) the latter can be obtained 
through $g_{xu}$ acting on an $(L-2)$-loop $x$-type color factor, 
and since $g_1$ commutes with $g_{xu}$, the same color factor
can obtained by the action of $g_{xu}$ on an $(L-1)$-loop
$x$-type color factor. 
\para

{\bf (Observation 2) All $L$-loop $x$-type color factors are obtained from
$x$-type color factors at $(L-1)$ and $(L-2)$ loops.}
To see this, observe that all $L$-loop $x$-type color factors are obtained
from $(L-1)$-loop color factors through the action of  $g_1$, $g_{ba}$, and $g_{ux}$.
However, from observation (1), any color factor obtained using $g_{ux}$
on an $(L-1)$-loop  $u$-type color factor can be obtained directly from an 
$(L-2)$-loop $x$-type color factor using $g_{xu} g_{ux}$.
This action typically produces a linear combination
of $x_a$- and $x_b$-type factors, 
so it will be useful to replace $g_{xu}g_{ux}$ with two-step operators 
(i.e., ones that map $(L-2)$-loop $x$-type color factors to $L$-loop $x$-type color factors)
that produce color factors of pure type: 
\begin{align}
g_{ab}^\Two
&=  \fr{1}{9} \left( g_{xu} g_{ux} - g_1^2 \right) \,,
\nn \\[2mm]
g_{bb}^\Two
&=  \fr{1}{18} \left( g_{xu} g_{ux} - 4 g_1^2 - 6 g_{ba} g_1 \right) \,.
\label{suntwostep}
\end{align}

These act on $x_a$- and $x_b$-type color factors respectively 
at $(L-2)$ loops to yield $x_b$-type color factors at $L$ loops
\begin{align} 
x_a g_{ab}^\Two &= x_b \,, \nn \\[2mm]
x_b g_{bb}^\Two&= x_b  \label{sungabtwo}
\end{align}
which are easily verified using eqs.~(\ref{sung1})-(\ref{sungux}).
Thus we have shown that all 
$L$-loop $x$-type color factors may be obtained from 
$x$-type color factors at $(L-1)$ and $(L-2)$ loops through
the action of the four operators  
$g_1$,  $g_{ba}$, $g_{ab}^\Two $, and $g_{bb}^\Two $.
\para

{\bf (Observation 3) All $L$-loop $x_a$-type color factors
can be obtained from $g_{ba}$ acting on an $(L-1)$-loop $x_b$-type 
color factor using \eqn{sungba} 
with one exception, namely $N^L x_a$, which results from $g_1$ 
acting repeatedly on the tree-level color factor $x_a$.}
From observation (2), all $x$-type color factors 
are obtained from $(L-1)$-loop $x$-type color
factors using $g_1$,  $g_{ba}$, $g_{ab}^\Two $, and $g_{bb}^\Two $,
but the last two always land on $x_b$-type color factors.
Because $g_1$ commutes with $g_{ba}$, 
the action of $g_1$ on an $(L-1)$-loop $x_a$-type color factor 
that is obtained from $g_{ba}$ acting on 
an $(L-2)$-loop $x_b$-type color factor
can also be obtained by $g_{ba}$ acting 
on an $(L-1)$-loop $x_b$-type color factor.
The remaining possibility is an $x_a$-type color factor obtained through 
$g_1$ acting $L$ times on the tree-level color factor  $x_a$. 
\para

{\bf (Observation 4)  
All $L$-loop $x_b$-type color factors for $L > 2$ can be obtained from
$g_1$ and $g_{bb}^\Two$ acting on 
$x_b$-type color factors at $L-1$ and $L-2$ loops.}
From observation (2), we know that 
all $L$-loop $x_b$-type color factors may be obtained 
from $g_1$ and $g_{bb}^\Two$ acting on $x_b$-type color factors at $L-1$
and $L-2$ loops respectively, 
and $g_{ab}^\Two$ acting on $x_a$-type color factors at $L-2$ loops.  
From observation (3), the latter may 
be replaced (with one possible exception discussed below)
by $g_{ba} g_{ab}^\Two$
acting on an $x_b$-type color factor at $L-3$ loops.
However, observe using \eqns{sungba}{sungabtwo} that
\begin{align}
x_b g_{ba} g_{ab}^\Two
= N x_a g_{ab}^\Two 
= N x_b
= x_b g_1 g_{bb}^\Two
\end{align}

thus the same color factor is obtained using $g_1$ and $g_{bb}^\Two$
alone.
The possible exception mentioned above 
is $g_{ab}^\Two$ acting on $N^{L-2} x_a$.
However since
\begin{align}
N^{L-2} x_a g_{ab}^\Two = 
x_a g_1^{L-2} g_{ab}^\Two = 
x_a g_{ab}^\Two g_1^{L-2} = 
x_b g_1^{L-2}
\end{align}
this is equivalent to $g_1$ acting repeatedly 
on the two-loop $x_b$-type color factor.
Thus, we have shown that all $x_b$-type color factors for $L>2$ 
can be generated by the action of two (commuting) operators 
\begin{align}
x_b g_1 		&= N x_b \,,\nn\\
x_b g_{bb}^\Two		&= x_b 
\end{align}
on lower-loop $x_b$-type color factors.
\para

From the observations above, we now determine the complete set
of $L$-loop color factors.
The most general $L$-loop $x_b$-type color factor is obtained by
acting with an arbitrary combination of $g_1$ and $g_{bb}^\Two$
on the two-loop color factor $x_b$ to give
\begin{align}
x_b g_1^{n_1} g_{bb}^{(2)  n_2}  
&= N^{n_1}  x_b 
\qquad \hbox{where} \qquad
n_1 + 2n_2 = L-2 
\label{sunxb}
\end{align}

where $n_1$ and $n_2$ are non-negative integers.
Thus, the space of $L$-loop $x_b$-type color factors is spanned by
$\lfloor L/2 \rfloor$ irreducible representations
\begin{align}
N^n x_b \,,
\qquad   0 \le n \le L-2, \qquad  n = L \mod 2  \,.
\label{sunspanxb} 
\end{align}
Using observation (3),
the space of $L$-loop $x_a$-type color factors (for $L \ge 1$) is spanned by
$\lfloor (L+1)/2 \rfloor$ irreducible representations
\begin{align}
N^n x_a \,,
\qquad   1 \le n \le L, \qquad  n = L \mod 2 \,.
\label{sunspanxa} 
\end{align}
Taking \eqns{sunspanxb}{sunspanxa} together, the number
of $L$-loop $x$-type irreducible representations (for $L \ge 1$) 
is given by $L$.  
\para

Using observation (1), 
the space of $L$-loop $u_a$-type color factors (for $L \ge 2$) 
is spanned by
$\lfloor L/2 \rfloor$ irreducible representations
\begin{align}
N^n u_a \,, 
\qquad 1 \le n \le L-1, \qquad  n = L-1 \mod 2
\label{sunspanua} 
\end{align}
and the space of $L$-loop $u_b$-type color factors 
is spanned by $\lfloor (L-1)/2 \rfloor$ irreducible representations
\begin{align}
N^n u_b \,,
\qquad 0 \le n \le L-3, \qquad  n = L-1 \mod 2  \,.
\label{sunspanub}  
\end{align}
Taking \eqns{sunspanua}{sunspanub} together, 
the number of $L$-loop $u$-type irreducible representations (for $L \ge 2$)
is given by $L-1$.
\para

\vfil\break

\begin{table}[t]
\begin{center}
\begin{tabular}{|l|*{7}{r}|c|}
\hline
\# of loops $L$   & 0 & 1 & 2  & 3  & 4  & 5  & 6  &  $L\ge 2$ \\
\hline
\# of $x_a$-type irreps & 1 & 1 & 1  & 2 & 2 & 3 & 3 &  
$ \lfloor (L+1)/ 2 \rfloor$ \\
\# of $x_b$-type irreps & 0 & 0 & 1  & 1 & 2 & 2 & 3 &  
$ \lfloor {L/ 2}\rfloor$ \\
\# of $u_a$-type irreps & 0 & 1 & 1  & 1 & 2 & 2 & 3  & 
$ \lfloor {L/ 2}\rfloor$ \\
\# of $u_b$-type irreps & 0 & 0 & 0  & 1 & 1 & 2 & 2  & 
$ \lfloor {(L-1)/ 2}\rfloor$ \\
\hline
total \# of color factors & 2 & 3 & 5  & 8 & 11& 14 & 17 & $3L-1$ \\
\hline
\end{tabular}
\end{center}
\caption{
Number of irreducible representations spanning the $L$-loop 
color space for \sun.} 
\label{table:suncolorspace}
\end{table}

Table \ref{table:suncolorspace}
summarizes the counting of irreducible representations 
spanning the $L$-loop color space for each value of $L$.
The total dimension of the $L$-loop color space given 
in the last row is the sum of these basis elements, 
taking into account that $x$-type elements 
are two-dimensional representations (of $S_4$)
while $u$-type elements are one-dimensional.
\para

\section{$L$-loop \sun\ null space}
\label{sec:sunnull}
\setcounter{equation}{0}

In the previous section, we generated  a complete set of
color factors spanning the $L$-loop color space for \sun,
which (for $L \ge 2$) is a $(3L-1)$-dimensional
subspace of the $3L+3$ dimensional trace space. 
In this section, we will determine the vectors that
span the $L$-loop null space, 
which is the four-dimensional orthogonal complement
to the $L$-loop color space. 
This will consist of two $u$-type null vectors
and one $x$-type irreducible representation. 
To do this, we first need to define
an inner project on the trace space.
\para

\subsection{Inner product}
\label{sec:inner}

To define an inner project, we need to represent 
color factors in a slightly different way.
Up to this point, we have represented a color factor
as a six-dimensional vector
\begin{align}
\CC =  
( \CC_{[1]},~ \CC_{[2]},~ \CC_{[3]};~ \CC_{[4]},~ \CC_{[5]},~ \CC_{[6]}) 
\label{innersix}
\end{align}

whose coefficients are polynomials in $N$.
We now express each of these polynomials as a vector.
An $L^{\rm th}$ degree polynomial 
may be written as an infinite-dimensional row vector 
\begin{align}
P(N) =\sum_{\ell=0}^{L}  P_\ell N^\ell 
\quad \to \quad 
 \bP = (P_0, P_1, P_2, \cdots, P_L, 0, \cdots )
\label{polyvector}
\end{align}
with all but the first $L+1$ entries of $\bP$ vanishing, that is
\begin{align}
\hbox{$P$ is an $L^{\rm th}$ degree polynomial} 
\quad\implies \quad
\bP = \bP \Pi_L
\quad \hbox{where} \quad 
\Pi_L = \begin{pmatrix}
	\unit_{(L+1)\times (L+1)} & 0 & \cdots \cr
	0 & 0 & \cdots \cr
\vdots & \vdots  & \ddots \cr
	\end{pmatrix}\,.
\label{defPi} 
\end{align}
We observe that for \sun, 
where the polynomials are of even or odd degree, 
every other entry of $\bP$ vanishes.
We may compactly express \eqn{polyvector} as 
\begin{align}
P(N) &= \bP  \bN^T, 
\qquad \hbox{where} \qquad 
\bN = (1, N, N^2, \cdots )\,.
\end{align}
Given two polynomials
$P = \bP  \bN^T $
and 
$P' =\bP' \bN^T $,
we may define a natural inner product $\langle P'| P \rangle$ by
\begin{align}
\langle P'| P \rangle =  \bP' \bP^T \,.
\end{align}

Extending this definition to color factors
(\ref{innersix})
we have
\begin{align}
\langle \CC'| \CC \rangle = \sumsix
 {\bf \CC'_{[\la]} } 
 {\bf \CC_{[\la]} }^T 
\end{align}

where 
$\CC_{[\lam]} = {\bf \CC_{[\la]} } \bN^T$.
If the color factor has the form
$\CC = [P, Q] \otimes v$ where $v=u$, $\xone$, or $\xtwo$, then the inner product become
\begin{align}
\langle \CC'| \CC \rangle = 
 \left( \bP'  \bP^T +\bQ' \bQ^T \right) 
\gamma_{v'v} 
\quad\hbox{where}\quad
\gamma = \begin{pmatrix}
3 & 0 & 0 \cr
0 & 2 & -1 \cr
0 & -1 & 2 \cr
\end{pmatrix} \,.
\end{align}

The main point here is that $u$-type color factors are orthogonal to $x$-type color factors.
Since we are only using the inner product to determine orthogonality, we will ignore the
$\gamma_{v'v}  $ piece and redefine 
\begin{align}
\langle \CC'| \CC \rangle = 
 \left( \bP' \bP^T +\bQ' \bQ^T \right) 
\delta_{v'v} 
\quad\hbox{where}\quad
v=u \hbox{ or } x\,.
\label{redefine}
\end{align}

Next we observe that the color factors 
are of the form 
$\CC =  c [p,q] \otimes v$, where 
$p$ and $q$ are (at most) degree-two polynomials,
$p = p_0 + p_1 N + p_2 N^2$ and $q = q_0 + q_1 N + q_2 N^2$,
and $c$ is a common factor of $P$ and $Q$.
Then the associated row vectors satisfy
\begin{align}
P = c p \implies \bP = \bc {\cal P} \quad\hbox{where}\quad
{\cal P} = 
\begin{pmatrix}
p_0 & p_1 & p_2 & 0   & \cdots \cr
  0 & p_0 & p_1 & p_2 & \cdots \cr
  0 &   0 & p_0 & p_1 & \cdots \cr
  0 &   0 &   0 & p_0 & \cdots \cr
\end{pmatrix} \,,
\nn\\
Q = c q \implies \bQ = \bc {\cal Q} \quad\hbox{where}\quad
{\cal Q} = 
\begin{pmatrix}
q_0 & q_1 & q_2 & 0   & \cdots \cr
  0 & q_0 & q_1 & q_2 & \cdots \cr
  0 &   0 & q_0 & q_1 & \cdots \cr
  0 &   0 &   0 & q_0 & \cdots \cr
\end{pmatrix} \,.
\label{PQmatrices}
\end{align}

The inner product (\ref{redefine}) between color factors
$\CC = c [p, q] \otimes v$ 
and 
$\CC' = c' [p', q'] \otimes v$ 
becomes 
\begin{align}
\langle \CC'| \CC \rangle = 
  \bc' M \bc^T \delta_{v'v} 
\quad \hbox{where} \quad
M = {\cal P' } {\cal P}^T +{\cal  Q' } {\cal Q}^T  \,.
\end{align}

We are interested in finding a set of null vectors $R$ which are orthogonal
to the color factors.  If $R$ has the form 
\begin{align}
R =  r [\tp, \tq ] \otimes v
\end{align}

then its inner product with a color factor $\CC = c [p, q] \otimes v'$ is 
\begin{align}
\langle \CC| R \rangle = 
  \bc M \br^T 
\delta_{v'v}
\quad \hbox{where} \quad
M = {\cal P } {\cal \tP}^T +{\cal  Q } {\cal \tQ}^T 
\label{innerCR}
\end{align}

where 
${\cal \tP} $ and  ${\cal \tQ}$
are defined analogously to \eqn{PQmatrices}.
An astute choice of $\tp$ and $\tq$
can ensure orthogonality of $R$ and $\CC$.
In particular, 
for degree one polynomials
$p = p_0 + p_1 N$ and $q= q_0 + q_1N$, 
we define 
$\tp = q_1 + q_0 N$ and $\tq = -  p_1 -p_0 N$
(possibly up to an overall sign for both).
For degree two polynomials 
$p = p_0 + p_1 N + p_2 N^2$ and $q= q_0 + q_1N + q_2N^2$, 
we define 
$\tp = q_2 +q_1 N+q_0 N^2$ and $\tq = -  p_2 -p_1 N - p_0 N^2$
(again possibly up to an overall sign).
Under these conditions,
one may easily verify that the matrix $M$ in \eqn{innerCR}
automatically vanishes, so that $\langle \CC| R \rangle = 0$.
This will be useful in defining the null space.
 
\subsection{\sun\ null vectors}

In sec.~\ref{sec:suncolor},
we determined a complete set of
color factors that span the $L$-loop color space for \sun,
namely,
\begin{align}
\CC^\Ell_{xa} &= c^\Ell_{xa} x_a,
& c^\Ell_{xa}& \in \{N^n \ | \  1 \le n \le L,\hskip10mm n = L \mod 2\} \,, \nn\\
\CC^\Ell_{xb} &= c^\Ell_{xb} x_b,
& c^\Ell_{xb}&\in \{N^n \  |\  0 \le n \le L-2,\hskip4mm n = L \mod 2\} \,,  \nn\\
\CC^\Ell_{ua} &= c^\Ell_{ua} u_a,
& c^\Ell_{ua}&\in \{N^n \ | \   1 \le n \le L-1,\hskip4mm n = L-1 \mod 2\} \,, \nn\\
\CC^\Ell_{ub} &= c^\Ell_{ub} u_b,
& c^\Ell_{ub}&\in \{N^n \ | \   0 \le n \le L-3,\hskip4mm n = L-1 \mod 2\} 
\label{suncolor}
\end{align}

where we recall that
\begin{align}
x_a &= [1,0]\otimes \xxi \,,  \nn \\
x_b &= [2,-N]\otimes \xxi \,, \nn \\
u_a &= [N,3]\otimes u \,,  \nn \\
u_b &= [N,N^2+3]\otimes u \,. 
\end{align}
In this subsection, we will obtain a complete 
set of $L$-loop null vectors $R^\Ell$,
defined to be orthogonal to the set (\ref{suncolor})
with respect to the inner project defined in the 
previous subsection.
We will show that the \sun\ null vectors can be 
of four possible types, namely,
\begin{align}
x_\alpha &= [0,1] \otimes \xxi \,, \nn\\
x_\beta &= [1,2N] \otimes \xxi \,, \nn\\
u_\alpha &= [3N,-1] \otimes \xxi \,, \nn\\
u_\beta &= [3N^2+1, -N] \otimes \xxi  \,.
\label{sunnull}
\end{align}
These are chosen,
using the prescription from the previous subsection,
so that 
$x_\alpha$-type null vectors are 
automatically orthogonal to $x_a$-type color factors,
$x_\beta$-type null vectors orthogonal to $x_b$-type color factors,
etc.
Also $x$-type null vectors are automatically orthogonal to $u$-type
color factors, and vice versa.
We use the remaining orthogonality condition to fully determine the 
form of the null vectors.
\para

\noindent{\bf (1) $x_\alpha$-type null vectors.} 
Consider an $L$-loop  null vector of the form
\begin{align}
R^\Ell_{x\alpha} = r^\Ell_{x\alpha} x_\alpha 
\end{align}
where $r^\Ell_{x\alpha}$ is a polynomial in $N$ of maximal degree $L-1$
and is odd/even for $L$ even/odd.
As we just remarked, 
orthogonality to $\CC^\Ell_{xa}$, $\CC^\Ell_{ua}$, and $\CC^\Ell_{ub}$
is automatic from the definition of $x_\alpha$.
To impose the final orthogonality condition, we compute
\begin{align}
\langle \CC^\Ell_{xb} | R^\Ell_{x\alpha} \rangle 
=  \bc^\Ell_{xb} M \br^{\Ell T}_{x\alpha}, \qquad 
M = \begin{pmatrix}
0 & -1 &  0 & \cdots \\
0 &  0 & -1 & \cdots \\
0 &  0 &  0 & \cdots \\
\vdots & \vdots & \vdots & \ddots \\
\end{pmatrix} \,.
\end{align}
where $M$ is defined by \eqn{innerCR}, 
using the ${\cal P } $ and ${\cal  Q }$
matrices appropriate to $x_b$ and $x_\alpha$.
Requiring $\langle \CC^\Ell_{xb} | R^\Ell_{x\alpha} \rangle  = 0$
for any $c^\Ell_{xb}$ belonging  to the set defined in \eqn{suncolor},
we find that $r^\Ell_{x\alpha}$ must vanish if $L$ is even, 
whereas for $L$ odd,  the only null vector is $R^\Ell_{x\alpha} = x_\alpha$.
\para

\noindent{\bf (2) $x_\beta$-type null vectors.} 
Next consider an $L$-loop  null vector of the form
\begin{align}
R^\Ell_{x\beta} = r^\Ell_{x\beta} x_\beta 
\end{align}
where $r^\Ell_{x\beta}$ is a polynomial in $N$  of maximal degree $L-2$
and is even/odd for $L$ even/odd.
Orthogonality to $\CC^\Ell_{xb}$, $\CC^\Ell_{ua}$, and $\CC^\Ell_{ub}$
is automatic from the definition of $x_\beta$.
To impose the final orthogonality condition, we compute
\begin{align}
\langle \CC^\Ell_{xa} | R^\Ell_{x\beta} \rangle 
=  \bc^\Ell_{xa} M \br^{\Ell T}_{x\beta}, \qquad 
M = \begin{pmatrix}
1 &  0 &  0 & \cdots \\
0 &  1 &  0 & \cdots \\
0 &  0 &  1 & \cdots \\
\vdots & \vdots & \vdots & \ddots \\
\end{pmatrix} \,.
\end{align}
where $M$ is defined by \eqn{innerCR}, 
using the ${\cal P } $ and ${\cal  Q }$
matrices appropriate to $x_a$ and $x_\beta$.
Requiring $\langle \CC^\Ell_{xa} | R^\Ell_{x\beta} \rangle  = 0$
for any $c^\Ell_{xa}$ belonging  to the set defined in \eqn{suncolor},
we see that $r^\Ell_{x\beta}$ must vanish if $L$ is odd, 
whereas for $L$ even, the only null vector is $R^\Ell_{x\beta} = x_\beta $.
\para

\noindent{\bf (3) $u_\alpha$-type null vectors.} 
Next consider an $L$-loop null vector of the form
\begin{align}
R^\Ell_{u\alpha} = r^\Ell_{u\alpha} u_\alpha 
\end{align}

where $r^\Ell_{u\alpha}$ is a polynomial in $N$  of maximal degree $L-1$
and is odd/even for $L$ even/odd.
Orthogonality to $\CC^\Ell_{xa}$, $\CC^\Ell_{xb}$, and $\CC^\Ell_{ua}$
is automatic from the definition of $u_\alpha$.
To impose the final orthogonality condition, we compute
\begin{align}
\langle \CC^\Ell_{ub} | R^\Ell_{u\alpha} \rangle 
=  \bc^\Ell_{ub} M \br^{\Ell T}_{u\alpha}, \qquad 
M = \begin{pmatrix}
0 &0 & -1 &  0 & \cdots \\
0 &0 &  0 & -1 & \cdots \\
0 &0 &  0 &  0 & \cdots \\
\vdots &\vdots & \vdots & \vdots & \ddots \\
\end{pmatrix} \,.
\end{align}
where $M$ is defined by \eqn{innerCR}, 
using the ${\cal P } $ and ${\cal  Q }$
matrices appropriate to $u_b$ and $u_\alpha$.
Requiring $\langle \CC_{ub} | R^\Ell_{u\alpha} \rangle  = 0$
for any $c^\Ell_{ub}$ belonging  to the set defined in \eqn{suncolor}
yields the null vector $R^\Ell_{u\alpha} = N u_\alpha $ for even $L$,
and $R^\Ell_{u\alpha} = u_\alpha$ for odd $L$.
\para

\noindent{\bf (4) $u_\beta$-type null vectors.} 
Finally consider an $L$-loop null vector of the form
\begin{align}
R^\Ell_{u\beta} = r^\Ell_{u\beta} u_\beta 
\end{align}

where $r_{u\beta}$ is a polynomial in $N$  of maximal degree $L-2$
and is even/odd for $L$ even/odd.
Orthogonality to $\CC^\Ell_{xa}$, $\CC^\Ell_{xb}$, and $\CC^\Ell_{ub}$
is automatic from the definition of $u_\beta$.
To impose the final orthogonality condition, we compute
\begin{align}
\langle \CC^\Ell_{ua} | R^\Ell_{u\beta} \rangle 
=  \bc^\Ell_{ua} M \br^{\Ell T}_{u\beta}, \qquad 
M = \begin{pmatrix}
0 & 1 &  0 &   \cdots \\
0 & 0 &  1 &   \cdots \\
0 & 0 &  0 &   \cdots \\
\vdots & \vdots & \ddots \\
\end{pmatrix} \,.
\end{align}
where $M$ is defined by \eqn{innerCR}, 
using the ${\cal P } $ and ${\cal  Q }$
matrices appropriate to $u_a$ and $u_\beta$.
Requiring $\langle \CC^\Ell_{ua} | R^\Ell_{u\beta} \rangle  = 0$
for any $c^\Ell_{ua}$ belonging  to the set defined in \eqn{suncolor},
yields the null vector $R^\Ell_{u\beta} = u_\beta $ for even $L$,
and $R^\Ell_{u\beta} = N u_\beta$ for odd $L$.
\para

\noindent{\bf Even-loop null space.}
To summarize the results of this section, 
the $L$-loop null space for even $L$ (with $L \ge 2$) 
is spanned by
\begin{align}
\hbox{Even loop ($L \ge 2$):}
\qquad
x_\beta, \qquad  N u_\alpha, \qquad u_\beta \,.
\end{align}

We may replace $u_\beta$ with 
$u_\beta - N u_\alpha = [1,0] \otimes u$,
and write the null vectors explicitly as
\begin{align}
\hbox{Even loop ($L \ge 2$):}
\qquad
[1, 2N] \otimes \xxi, \qquad 
[3N^2,-N] \otimes u , \qquad 
[1,0] \otimes u \,.
\label{evenloopnull}
\end{align}
At tree level, the only null vector is $[1,0] \otimes u$.
\para

\noindent{\bf Odd-loop null space.}
The  $L$-loop null space for odd $L$ (with $L \ge 3$) is spanned by
\begin{align}
\hbox{Odd loop ($L \ge 3$):}
\qquad
x_\alpha, \qquad   u_\alpha, \qquad N u_\beta \,.
\end{align}

Writing the null vectors explicitly, we have
\begin{align}
\hbox{Odd loop ($L \ge 3$):}
\qquad
[0,1]\otimes \xxi, \qquad   [3N,-1] \otimes u, 
\qquad [3N^3 + N, -N^2] \otimes u \,.
\label{oddloopnull}
\end{align}

At one loop, the null vectors are 
$ [0,1]\otimes \xxi$ and $  [3N,-1] \otimes u$.
\para

\Eqns{evenloopnull}{oddloopnull} agree precisely 
with the results obtained in ref.~\cite{Naculich:2011ep,Edison:2012fn},
taking into account footnote \ref{fn1}.
Thus, 
as stated in the introduction,
for \sun \ there are precisely four $L$-loop null vectors for all $L \ge 2$.

\section{$L$-loop \son\ color space}
\label{sec:soncolor}
\setcounter{equation}{0}

The goal of this section is to explicitly construct the 
space of $L$-loop color factors for \son.
The procedure is analogous to that employed 
in sec.~\ref{sec:suncolor}.
An $L$-loop color factor may be expressed as
\begin{align}
[P, Q] \otimes u, \qquad\qquad  [P, Q] \otimes \xxi \qquad (i=1,2)
\end{align}
where $P$ and $Q$ are polynomials in $N$
of maximal degree $L$  and $L-1$ respectively. 
Thus, $L$-loop \son\ color factors inhabit a vector space $V^\Ell$ 
of dimension $6L+3$ (the trace space). 
The polynomials $P$ and $Q$
corresponding to color factors, however,
are not completely arbitrary but satisfy certain constraints. 
Consequently, the space of all $L$-loop color factors 
spans a proper subspace (the color space) of $V^\Ell$.
\para

As before,
we iteratively construct an explicit basis 
for the $L$-loop \son\ color space,
beginning with the 
single tree-level irreducible representation $[1,0]\otimes \xxi$
and acting repeatedly with the iterative matrices for \son\ obtained
in sec.~\ref{sec:iterative}
\begin{align}
g_1 &= \begin{pmatrix} K & 0 \cr 0 & K \end{pmatrix}, \quad
g_{xx} = \begin{pmatrix} K & 0 \cr -4 & 0 \end{pmatrix}, \quad
g_{xu} = \begin{pmatrix} K-6 & 3 \cr 0 & -2K \end{pmatrix}, \quad
g_{ux} = \begin{pmatrix} K-3 & -3 \cr 12 & -2K \end{pmatrix}
\label{gson}
\end{align}

where we have chosen to set $e=2$.
Moreover, we find it convenient to express these matrices 
in terms of the \son\ quadratic Casimir 
$K = N-2 $ rather than in terms of $N$.
As before,
these $2 \times 2$ matrices act on the $[P,Q]$ part of the color factor,
while the subscripts indicate their  action on the 
$x$ or $u$ part of the color factor,
so that, for example,
$g_{xu}$ takes an $x$-type color factor to a $u$-type color factor,
etc.
We will show that these matrices generate a basis
consisting of polynomials multiplied by
one of four specific (linearly independent) types: 
\begin{align}
x_a      & \equiv [1,0]\otimes \xxi,  \nn \\[1mm]
x_b      & \equiv [K+3,1]\otimes \xxi, \nn \\[1mm]
u_a      & \equiv [K-6,3]\otimes u,  \nn \\[1mm]
u_b      & \equiv [(K+3)(K-6),K+9]\otimes u \,.
\label{sontypes}
\end{align}

Our first step is to ascertain how each of the operators (\ref{gson})
act on the types of color factors (\ref{sontypes}).
First, the operator $g_1$ just rescales each type by $K$
\begin{align}
x_a  g_1 &= K x_a \,,\nn\\
x_b  g_1 &= K x_b \,,\nn\\
u_a  g_1 &= K u_a \,,\nn\\
u_b  g_1 &= K u_b  \,.
\label{song1}
\end{align}

Second, the operator $g_{xx}$ acts on the $x$-type color factors as
\begin{align}
x_a  g_{xx} &= K x_a \,,\nn\\
x_b  g_{xx} &= (K+4)(K-1) x_a  \,.
\end{align}
Since the action of $g_{xx}$ on $x_a$ is identical to the action of $g_1$
(and therefore redundant), 
we will restrict our attention to its action on $x_b$, defining $g_{ba} = g_{xx}$ with
\begin{align}
x_b g_{ba} &= (K+4)(K-1) x_a  \,.
\label{songba}
\end{align}

Third, the operator $g_{xu}$ takes an $x_a$-type color factor 
to a $u_a$-type color factor,
and  an $x_b$-type color factor to a $u_b$-type color factor:
\begin{align}
x_a g_{xu} &= u_a  \,,\nn \\
x_b g_{xu} &= u_b  \,.
\label{songxu}
\end{align}

Finally, the operator $g_{ux}$ acts on $u$-type color factors to give 
linear combinations of $x_a$ and $x_b$ types: 
\begin{align}
u_a g_{ux} &= 
10K^2 \, x_a   - 9 (K-2) x_b
\,,\nn \\
u_b g_{ux} &= 
6K(K-1)(K+4) x_a  
-(5K^2+9K-54) x_b
\,.
\label{songux}
\end{align}

Our next step is to generate the \son\ color space 
through three-loop order. 
We begin with 
the single tree-level irreducible representation
\begin{align}
\hbox{Tree level:}\qquad     x_a \,.
\label{sontreelevel}
\end{align}

Acting on $x_a$ with $g_1$ using \eqn{song1} 
and with $g_{xu}$ using \eqn{songxu}, 
we obtain the three-dimensional space spanned by
two irreducible representations
\begin{align}
\hbox{One loop:}\qquad  K  x_a, \quad  u_a \,.
\label{sononeloop}
\end{align}

We then act on each of these one-loop color factors 
with $g_1$ to obtain $K^2 x_a $ and $K u_a$. 
The action of $g_{xu}$ on $K x_a$ is redundant,
but we can act on the $u_a$-type color factor 
with $g_{ux}$ to obtain $[K^2-9K+54,-9K+18] \otimes \xxi$,
which is a linear combination of $x_a$ and $x_b$ types,
as shown in \eqn{songux}.
Since we already have $K^2 x_a$ in the color space, we subtract 
$10 K^2 x_a$ and divide by $-9$ to obtain $(K-2) x_b$.  
Thus the two-loop color space is five-dimensional,
spanned by
three irreducible representations
\begin{align}
\hbox{Two loops:}\qquad  
K^2 x_a \,, 
\quad K u_a \,,
\quad (K-2) x_b \,.
\label{sontwoloop}
\end{align}

Observe that all these results are consistent with the results obtained 
earlier in \eqn{sontreeonetwo},
noting that $[N-8,N-4]\otimes \xxi$ is a linear combination of 
$K^2 x_a$  and $(K-2) x_b$.
The three-loop color factors are then obtained by acting 
on each of the two-loop color factors with $g_1$. 
The action of $g_{ux}$ on $K u_a$ is redundant.
We can also act on $(K-2) x_b$
with $g_{ba}$ using \eqn{songba}  and with $g_{xu}$ using \eqn{songxu}.
The three-loop color space is thus eight-dimensional, spanned by 
five irreducible representations
\begin{align}
\hbox{Three loops:}\qquad  
K^3 x_a\,, 
\quad (K+4)(K-1)(K-2) x_a\,,
\quad K^2 u_a \,,
\quad K(K-2) x_b\,, 
\quad (K-2) u_b \,.
\label{sonthreeloop}
\end{align}

We now make some general observations
that allow us to determine the complete span of color factors
at arbitrary loop order $L$.
We omit the arguments for these when they are identical to those
given for \sun\ in sec.~\ref{sec:suncolor}.
\para

{\bf (Observation 1) All $L$-loop  $u$-type color factors are generated 
by the action of $g_{xu}$ on the complete set of $x$-type 
color factors at $(L-1)$ loops using \eqn{songxu}.}
\para

{\bf (Observation 2) All $L$-loop $x$-type color factors are obtained from
$x$-type color factors at $(L-1)$ and $(L-2)$ loops.}
The two-step operators that produce color factors of pure type are 
\begin{align}
g_{ab}^\Two
&=  \fr{1}{9} \left( - g_{xu} g_{ux} + 10 g_1^2 \right) \,,
\nn \\[2mm]
g_{bb}^\Two
&=  \fr{1}{9} \left( - g_{xu} g_{ux} - 5 g_1^2 + 6 g_{ba} g_1 \right) 
\label{sontwostep}
\end{align}

which  act on $x_a$- and $x_b$-type color factors respectively 
at $(L-2)$ loops to yield $x_b$-type color factors at $L$ loops
\begin{align} 
x_a g_{ab}^\Two &= (K-2) x_b \,, \nn \\[2mm]
x_b g_{bb}^\Two&= (K-6) x_b  
\end{align}
easily verified using eqs.~(\ref{song1})-(\ref{songux}).
Thus all $L$-loop $x$-type color factors may be obtained from 
$x$-type color factors at $(L-1)$ and $(L-2)$ loops through
the action of the four operators  
$g_1$,  $g_{ba}$, $g_{ab}^\Two $, and $g_{bb}^\Two $.
\para

{\bf (Observation 3) All $L$-loop $x_a$-type color factors can be obtained 
from $g_{ba}$ acting on an $(L-1)$-loop $x_b$-type color factor
using \eqn{songba} 
with one exception, namely $K^L x_a$, which results from $g_1$ 
acting repeatedly on the tree-level color factor $x_a$.}
\para

{\bf (Observation 4)  
All $L$-loop $x_b$-type color factors for $L > 2$ can be obtained from
$g_1$, $g_{bb}^\Two$, and $g_{bb}^\Three$ acting on
$x_b$-type color factors at $L-1$, $L-2$, and $L-3$ loops.}
From observation (2), we know that all $L$-loop $x_b$-type 
color factors may be obtained from 
$g_1$ and $g_{bb}^\Two$ acting on $x_b$-type color factors at $L-1$
and $L-2$ loops respectively, 
and $g_{ab}^\Two$ acting on $x_a$-type color factors at $L-2$ loops.  
From observation (3), 
the latter may be replaced (with one possible exception)
by $g_{ba} g_{ab}^\Two$
acting on an $x_b$-type color factor at $L-3$ loops.
The one possible exception is $g_{ab}^\Two$ acting on $K^{L-2} x_a$.
However since
\begin{align}
K^{L-2} x_a g_{ab}^\Two = 
x_a g_1^{L-2} g_{ab}^\Two = 
x_a g_{ab}^\Two g_1^{L-2} = 
(K-2) x_b g_1^{L-2}
\end{align}
this is equivalent to $g_1$ acting repeatedly on the two-loop $x_b$-type color factor.
It is convenient to replace 
$g_{ba} g_{ab}^\Two$
with a three-step operator 
\begin{align}
g_{bb}^\Three 
&=  \fr{1}{4} \left[ -g_{ba} g_{ab}^\Two  + g_1^3 + g_1 g_{bb}^\Two \right]
=  \fr{1}{36} \left[ g_{ba} g_{xu} g_{ux} - g_1 g_{xu} g_{ux} 
	- 4 g_{ab} g_1^2 + 4 g_1^2 \right]
\end{align}
which maps an $(L-3)$-loop $x_b$-type color factor 
to an $L$-loop $x_b$-type color factor
\begin{align}
x_b g_{bb}^\Three&= (K-2) x_b \,.
\end{align}

To summarize, all $x_b$-type colors for $L>2$ can be generated by
the action of three (commuting) operators 
\begin{align}
x_b g_1 		&= K x_b \,, \nn\\
x_b g_{bb}^\Two		&= (K-6) x_b  \,, \nn\\
x_b g_{bb}^\Three 	&= (K-2) x_b
\label{threeg}
\end{align}
acting on lower-loop $x_b$-type color factors.
\para

From the observations above, 
we are now able to determine the complete set of
$L$-loop color factors.  
Beginning with $x_b$-type color factors, 
we observe from \eqn{sontwoloop}
that the first $x_b$-type color factor occurs at two loops,
namely $(K-2) x_b$.
The set of all higher-loop $x_b$-type color factors
is obtained by acting on $(K-2) x_b$ with an arbitrary combination of 
$g_1$, $g_{bb}^\Two$, and $g_{bb}^\Three$:
\begin{align}
(K-2) x_b g_1^{n_1} g_{bb}^{(2)  n_2}  g_{bb}^{(3)  n_3} 
&= K^{n_1} (K-6)^{n_2} (K-2)^{n_3+1} x_b 
\label{sonxb}
\end{align}

where $n_1$, $n_2$, and $n_3$ are arbitrary non-negative integers
that satisfy
\begin{align}
n_1 + 2n_2 + 3 n_3 = L-2 \,.
\label{integercondition}
\end{align}

The right hand side of \eqn{sonxb} may be written more explicitly as 
\begin{align}
[P, Q] \otimes \xxi
\qquad \hbox{with} \begin{cases}
P = K^{n_1} (K-6)^{n_2} (K-2)^{n_3+1} (K+3) \\
Q = K^{n_1} (K-6)^{n_2} (K-2)^{n_3+1}
\end{cases}
\end{align}
confirming that $P$ is a polynomial of maximal degree $L$ and
$Q$ is a polynomial of maximal degree $L-1$.
For $L=3$ through $L=7$, the number of solutions of 
\eqn{integercondition} is $L-2$, 
with $n_3$ given by either 0 or 1.
Specifically, denoting $n_2=n$ and $n_1 = L-2-2n-3n_3$,
these solutions correspond to $x_b$-type irreducible representations
\begin{align}
&K^{L-2-2n} (K-6)^{n} (K-2) x_b, \qquad n=0, \cdots, \lfloor { L-2 \over 2 }\rfloor \,, \nn\\ 
&K^{L-5-2n} (K-6)^{n} (K-2)^2 x_b, \qquad n=0, \cdots, \lfloor { L-5 \over 2 }\rfloor \,.  
\label{listsonxb}
\end{align}

This set of $L-2$ irreducible representations 
(for $L \ge 3$) is linearly independent, 
since the exponents of $K$ are all distinct.
Starting at $L=8$, 
additional solutions of \eqn{integercondition} arise, 
with $n_3 \ge 2$,
but we claim that the corresponding color factors 
are not linearly independent of the set (\ref{listsonxb}).
We verify this claim in appendix \ref{sec:appB},
where we explicitly construct two 
$x_\alpha$-type irreducible representations 
orthogonal to the entire set (\ref{sonxb}).
This establishes that at most 
$L-2$ of the irreducible representations 
in \eqn{sonxb} are independent.
With $L-2$ as both lower and upper bound, 
the $L-2$ irreducible representations  belonging to the set (\ref{listsonxb}) 
constitute a complete and independent set of $L$-loop $x_b$-type color factors for $L \ge 3$.
From these, we may construct the rest of the color space 
using the observations above.
\para

Let us first consider the $x_a$-type color factors.
From \eqn{sontreelevel} through \eqn{sonthreeloop},
we observe that there is 
one $x_a$-type irreducible representation for $L=0$ through $L=2$,
and two $x_a$-type irreducible representations for $L=3$.
For $L \ge 3$, we can use observation (3) 
to generate a complete set of linearly independent 
$L$-loop $x_a$-type color factors by acting
with $g_{ba}$ on the complete set of $(L-1)$-loop $x_b$-type  color factors 
in \eqn{listsonxb}
and adding in the one exception:
\begin{align}
&K^L x_a    \,, \nn\\
&K^{L-3-2n} (K-6)^{n} (K-2) (K-1) (K+4)  x_a, \qquad n=0, \cdots, \lfloor { L-3 \over 2 }\rfloor   \,, \nn\\ 
&K^{L-6-2n} (K-6)^{n} (K-2)^2 (K-1) (K+4) x_a, \qquad n=0, \cdots, \lfloor { L-6 \over 2 }\rfloor
\,.
\label{listsonxa}
\end{align}

This set contains (for $L \ge 4$) $L-2$ linearly independent $x_a$-type 
irreducible representations.
\para

Next we turn to $u$-type color factors.
From \eqn{sonthreeloop}, we observe that 
the first $u_b$-type color factor occurs at $L=3$, 
namely, $(K-2) u_b$.
From observation (1) above, 
we can generate all $L$-loop $u_b$-type color factors for $L \ge 3$ 
by acting with $g_{xu}$ 
on the complete set of $(L-1)$-loop $x_b$-type color factors 
in \eqn{listsonxb} to give
\begin{align}
&K^{L-3-2n} (K-6)^{n} (K-2) u_b, \qquad n=0, \cdots, \lfloor { L-3 \over 2 }\rfloor \,, \nn\\ 
&K^{L-6-2n} (K-6)^{n} (K-2)^2 u_b, \qquad n=0, \cdots, \lfloor { L-6 \over 2 }\rfloor 
\label{listsonub}
\end{align}

which is a complete set of (for $L \ge 4$) 
$L-3$ $u_b$-type color factors at $L$ loops.
Finally, we observe that there is one $u_a$-type color factor at $L=1$ through $L=3$,
and two $u_a$-type color factors at $L=4$.
Using observation (3) above, 
we can generate
all $L$-loop $u_a$-type color factors for $L \ge 4$ by acting with $g_{xu}$ 
on the complete set of $(L-1)$-loop $x_a$-type  
color factors in \eqn{listsonxa} to obtain
\begin{align}
&K^{L-1} u_a \,, \nn\\
&K^{L-4-2n} (K-6)^{n} (K-2) (K-1) (K+4)  u_a, \qquad n=0, \cdots, \lfloor { L-4 \over 2 }\rfloor \,, \nn\\ 
&K^{L-7-2n} (K-6)^{n} (K-2)^2 (K-1) (K+4) u_a, \qquad n=0, \cdots, \lfloor { L-7 \over 2 }\rfloor
\label{listsonua}
\end{align}

which is a complete set of (for $L \ge 5$) 
$L-3$ $u_a$-type color factors at $L$ loops.
\para

We summarize the counting of irreducible representations 
spanning the $L$-loop color space in table \ref{table:soncolorspace}.
The total dimension of the $L$-loop color space given 
in the last row is the sum of these basis elements, 
taking into account that $x$-type elements 
are two-dimensional representations (of $S_4$)
while $u$-type elements are one-dimensional.
Observe that the dimensions of the color spaces begins
to differ from those of \sun\ at $L=5$.
\para

\begin{table}[t]
\begin{center}
\begin{tabular}{|l|*{7}{r}|c|}
\hline
\# of loops $L$   & 0 & 1 & 2  & 3  & 4  & 5  & 6  &  $L\ge 5$ \\
\hline
\# of $x_a$-type irreps & 1 & 1 & 1  & 2 & 2 & 3 & 4 &  $L-2$ \\
\# of $x_b$-type irreps & 0 & 0 & 1  & 1 & 2 & 3 & 4 &  $L-2$ \\
\# of $u_a$-type irreps & 0 & 1 & 1  & 1 & 2 & 2 & 3  & $L-3$\\
\# of $u_b$-type irreps & 0 & 0 & 0  & 1 & 1 & 2 & 3  & $L-3$\\
\hline
total \# of color factors & 2 & 3 & 5  & 8 & 11& 16& 22 & $6L-14$ \\
\hline
\end{tabular}
\end{center}
\caption{
Number of irreducible representations spanning the $L$-loop color 
space for \son.}
\label{table:soncolorspace}
\end{table}

\section{$L$-loop \son\ null space}
\label{sec:sonnull}
\setcounter{equation}{0}

In the previous section, we generated a basis of 
color factors spanning the $L$-loop color space for \son.
The numbers of independent color factors for various values of $L$
are given in table \ref{table:son}, divided into $x$-type and 
$u$-type.\footnote{The number of $x$-type color factors is twice
the number of $x$-type irreducible representations because
those representations are two-dimensional.}
The dimensions of the associated trace spaces, 
in which these color factors live, are also listed by type.
The differences of these two numbers is the number of null vectors, 
defined as inhabiting the orthogonal complement to the color space,
and are also listed in the table. 
The last three rows of the table,
which list the dimensions of the spaces regardless of type,
reproduce table \ref{table:sonnull}.
\para

\begin{table}[t]
\begin{center}
\begin{tabular}{|l|*{7}{r}|c|}
\hline
number of loops           & 0 & 1 & 2  & 3  & 4  & 5  & 6  &  $L\ge 5$ \\
\hline
\# $x$-type color factors & 2 & 2 & 4  & 6  &  8 & 12 & 16 & $4L-8$\\
dim $x$-type trace space  & 2 & 6 & 10 & 14 & 18 & 22 & 26 & $4L+2$ \\
\# $x$-type null vectors  & 0 & 4 & 6  & 8  & 10 & 10 & 10 & 10 \\
\hline
\# $u$-type color factors & 0 & 1 & 1  & 2  & 3  & 4  & 6 &  $2L-6$\\
dim $u$-type trace space  & 1 & 3 & 5  & 7  & 9  & 11 & 13 & $2L+1$ \\
\# $u$-type null vectors  & 1 & 2 & 4  & 5  & 6  & 7  & 7 & 7 \\
\hline
\# color factors          & 2 & 3 & 5  & 8  &11  & 16 & 22&  $6L-14$\\
dim trace space           & 3 & 9 &15  &21  &27  & 33 & 39 & $6L+3$ \\
\# null vectors           & 1 & 6 &10  &13  &16  & 17 & 17& 17\\
\hline
\end{tabular}
\end{center}
\caption{
Dimensions of trace, color, and null spaces for \son\ amplitudes.}
\label{table:son}
\end{table}

For $L\ge 5$,
the color space is a $(6L-14)$-dimensional
subspace of the $(6L+3)$-dimensional trace space,
so the null space is therefore (as claimed in the introduction)
generically 17-dimensional,
and consists of 10 $x$-type null vectors
and 7 $u$-type null vectors. 
\para

The purpose of this section is to derive explicit expressions 
for the 10 $x$-type null vectors.
(The construction of the 7 $u$-type null vectors is left to future work.)
As before, we first need to define an inner project on the trace space.
\para

\subsection{Inner product}

We choose an inner project for the \son\ trace space 
similar to that defined in sec.~\ref{sec:inner} for the \sun\ trace space,
but with a slight difference.
The inner product of two polynomials $P$ and $P'$
is given by
\begin{align}
\langle P'| P \rangle =  \bP' \bP^T
\end{align}

except that the row vectors
$ \bP = (P_0, P_1, P_2, \cdots )$
consist of the coefficients of the polynomials expressed 
in terms of $K=N-2$ rather than of $N$: 
\begin{align}
P(K) &= \bP  \bK^T, 
\qquad \hbox{with} \qquad 
\bK = (1, K, K^2, \cdots ) \,.
\end{align}

Other than this, everything is the same as in sec.~\ref{sec:inner}.

\subsection{\son\ null vectors}

In sec.~\ref{sec:soncolor},
we determined a complete set of
color factors that span the $L$-loop color space for \son,
namely,
\begin{align}
\CC^\Ell_{xa} &= c^\Ell_{xa} x_a,
& c^\Ell_{xa}& 
\in  \{ K^L \} \cup \{ (K-1)(K+4) c^{(L-1)}_{xb} \} \,, \nn\\
\CC^\Ell_{xb} &= c^\Ell_{xb} x_b,
& c^\Ell_{xb}&
\in \{ K^{n_1} (K-6)^{n_2} (K-2)^{n_3+1} \ | \ n_1 + 2n_2 + 3 n_3 = L-2 \} \,, \nn\\
\CC^\Ell_{ua} &= c^\Ell_{ua} u_a,
& c^\Ell_{ua}& \in \{ c^{(L-1)}_{xa}\}  \,, \nn\\
\CC^\Ell_{ub} &= c^\Ell_{ub} u_b,
& c^\Ell_{ub}& \in \{ c^{(L-1)}_{xb} \} 
\label{soncolor}
\end{align}

where we recall that
\begin{align}
x_a      & = [1,0]\otimes \xxi,  \nn \\
x_b      & = [K+3,1]\otimes \xxi, \nn \\
u_a      & = [K-6,3]\otimes u,  \nn \\
u_b      & = [(K+3)(K-6),K+9]\otimes u \,.
\end{align}

We will obtain a complete set of $x$-type null vectors $R_x^\Ell$
living in the $L$-loop trace space and orthogonal to the set (\ref{soncolor}).
We will show that all \son\ $x$-type null vectors can
be written in terms of 
three  possible types, namely,
\begin{align}
x_\alpha &= [0,1] \otimes \xxi \,,\nn\\
x_\beta &= [K ,-3K-1] \otimes \xxi \,,\nn\\
x_\gamma &= [1,0] \otimes \xxi \,.
\label{sonnull}
\end{align}

These are chosen,
using the prescription given at the end of sec.~\ref{sec:inner},
so that 
$x_\alpha$-type null vectors are 
automatically orthogonal to $x_a$-type color factors
and the 
$x_\beta$-type null vectors are orthogonal to $x_b$-type color factors.
Unlike \sun, we will also need a third type, $x_\gamma$,  of null vector
which is not automatically 
orthogonal to either $x_a$- or $x_b$-type color factors.
All $x$-type null vectors, however, 
are automatically orthogonal to the $u$-type color factors.
The ten $x$-type null vectors (for $L \ge 5$) 
consist of five $x$-type irreducible representations:
two each of $x_\alpha$ and $x_\beta$ type,
and one of $x_\gamma$ type.
\para

\noindent{\bf (1) $x_\alpha$-type null vectors.} 
Consider an $L$-loop  null vector of the form
\begin{align}
R^\Ell_{x\alpha} = r^\Ell_{x\alpha} x_\alpha 
\end{align}
where $r^\Ell_{x\alpha}$ is a polynomial in $K$ of maximal degree $L-1$.
Orthogonality to $\CC^\Ell_{xa}$, $\CC^\Ell_{ua}$, and $\CC^\Ell_{ub}$
is automatic from the definition of $x_\alpha$.
The final orthogonality condition gives 
\begin{align}
0= \langle \CC^\Ell_{xb} | R^\Ell_{x\alpha} \rangle 
=  \bc^\Ell_{xb} M_{b\alpha} \br^{\Ell T}_{x\alpha}, \qquad 
M_{b\alpha} = \begin{pmatrix}
1 &  0 &  0 & \cdots \\
0 &  1 &  0 & \cdots \\
0 &  0 &  1 & \cdots \\
\vdots & \vdots & \vdots & \ddots \\
\end{pmatrix}
\label{sonbalpha}
\end{align}
where $M_{b\alpha}$ is defined in \eqn{innerCR} using the 
${\cal P } $ and ${\cal  Q }$
matrices appropriate to $x_b$ and $x_\alpha$.
\para

Let's now impose
$\langle \CC^\Ell_{xb} | R^\Ell_{x\alpha} \rangle =0$ to determine
the form of the $x_\alpha$-type null vectors.
At one loop, this condition is automatically satisfied since there are
no one-loop $x_b$-type color factors, so we have
\begin{align}
R^\One_{x\alpha} =  x_\alpha \,.
\end{align}

At two loops, we use $c_{xb}^\Two = K-2$ to find
\begin{align}
R^\Two_{x\alpha} =  (K+ \half) x_\alpha \,.
\label{twoloopnullxalpha}
\end{align}

At three loops, we use 
$c_{xb}^\Three = K(K-2)$ to find
$ R^\Three_{x\alpha} =  (K^2+ \half K + \lambda) x_\alpha $
where $\lambda$ is arbitrary.   
Thus the null space contains
a pair of independent $x_\alpha$-type irreducible representations, 
which we choose to be (for reasons that will immediately become clear)
\begin{align}
R^\Three_{x\alpha,1} &=  (K^2+ \half K + \fr{1}{4} ) x_\alpha \,, \nn\\
R^\Three_{x\alpha,2} &=  (K^2+ \half K + \fr{5}{36} ) x_\alpha \,.
\label{threeloopnullxalpha}
\end{align}

In appendix \ref{sec:appB}, we prove that, for all $L \ge 3$ 
there are exactly two $x_\alpha$-type irreducible representations,
given by 
\begin{align}
R^\Ell_{x\alpha,j} &=  r^\Ell_{x\alpha,j} x_\alpha \,, \qquad  j=1,2 \nn\\
r^\Ell_{x\alpha,1} &=  
\frac{\left[ 1 - (2K)^L \right] }{2^{L-1} (1-2K)} 
\,,\nn\\
r^\Ell_{x\alpha,2} &=  
\frac{4\left[ 1 - (3K)^L \right] }{3^{L} (1-3K)} 
+
\frac{2\left[ 1 - (-6K)^L \right] }{(-6)^{L} (1+6K)} 
\,.
\label{twoxalphasolns}
\end{align}

At three loops, these agree with 
\eqn{threeloopnullxalpha}, 
and at four loops, they give 
\begin{align}
r^\Four_{x\alpha,1} &=  (K^3+ \half K^2 + \fr{1}{4}K + \fr{1}{8} ) \,, \nn\\
r^\Four_{x\alpha,2} &=  (K^3+ \half K^2 + \fr{5}{36}K + \fr{11}{216} ) \,.
\end{align}

There is an evident pattern whereby
$r^\Ell_{x\alpha,j}$ is given by $K r^{(L-1)}_{x\alpha,j}$
plus a constant easily obtained from \eqn{twoxalphasolns}.
\para

\noindent{\bf (2) $x_\beta$-type null vectors.} 
Next consider an $L$-loop  null vector of the form
\begin{align}
R^\Ell_{x\beta} = r^\Ell_{x\beta} x_\beta 
\end{align}
where $r^\Ell_{x\beta}$ is a polynomial in $K$ of maximal degree $L-2$.
Orthogonality to $\CC^\Ell_{xb}$, $\CC^\Ell_{ua}$, and $\CC^\Ell_{ub}$
is automatic from the definition of $x_\beta$.
The final orthogonality condition gives
\begin{align}
0= \langle \CC^\Ell_{xa} | R^\Ell_{x\beta} \rangle 
=  \bc^\Ell_{xa} M_{a\beta} \br^{\Ell T}_{x\beta}, \qquad 
M_{a\beta} = \begin{pmatrix}
0 & 0 & 0  & \cdots \\
1 & 0 & 0  & \cdots \\
0 &1  & 0  & \cdots \\
\vdots & \vdots & \vdots & \ddots \\
\end{pmatrix} 
\label{sonabeta}
\end{align}

where $M_{a\beta}$ is defined in \eqn{innerCR} using the 
${\cal P } $ and ${\cal  Q }$
matrices appropriate to $x_a$ and $x_\beta$.
One of the $L$-loop $x_a$-type color factors is $c^\Ell_{xa} = K^L$,
for which \eqn{sonabeta} is automatically satisfied since 
$r_{x\beta}$ has maximal degree $L-2$.
By observation (3) of sec.~\ref{sec:soncolor},
all the other $L$-loop $x_a$-type color factors are obtained from
$(L-1)$-loop $x_b$-type color factors,
\begin{align}
c^\Ell_{xa} =  (K-1)(K+4) c^{(L-1)}_{xb}
\end{align}

which can be expressed in matrix form as 
\begin{align}
\bc^\Ell_{xa} &=   \bc^{(L-1)}_{xb} G_{ba}, 
&
G_{ba} &=
\begin{pmatrix}
-4 &3 & 1 &  0 & \cdots \\
0 &-4 & 3 &  1 & \cdots \\
0 &0 & -4  & 3 & \cdots \\
\vdots &\vdots & \vdots & \vdots & \ddots \\
\end{pmatrix} \,.
\label{defGba}
\end{align}

Thus \eqn{sonabeta} implies
\begin{align}
0& =
\bc^{(L-1)}_{xb}  H \br^{\Ell T}_{x\beta}, &
H &= G_{ba} M_{a\beta} = \begin{pmatrix}
3 & 1 & 0 & 0 & \cdots \\
-4& 3 & 1 & 0 & \cdots \\
0 &-4 & 3 & 1 & \cdots \\
0 & 0 &-4 & 3 & \cdots \\
\vdots & \vdots & \vdots & \ddots \\
\end{pmatrix} \,.
\label{defH}
\end{align}

Given that $c^{(L-1)}_{xb}$  and $r^{(L)}_{x\beta}$ 
are both of maximal degree $L-2$,
we may truncate the infinite matrix $H$ 
to the finite matrix $H^\Ell$,
consisting of 
the first $L-1$ rows and columns of $H$.
Then \eqn{defH} becomes
\begin{align}
0 &=
\bc^{(L-1)}_{xb}  H^{(L)} \br^{\Ell T}_{x\beta},
& H^{(L)} &= \Pi_{L-2} H \Pi_{L-2} \,.
\end{align}

We observe\footnote{
This follows from 
$\det H^\Ell = 3 \det H^{(L-1)} + 4 \det H^{(L-2)}$.}
that $\det H^\Ell > 0$, 
so that the $(L-1)\times (L-1)$ matrix $H^\Ell$ is
invertible.
Since generically we found two solutions to 
$\bc^{(L-1)}_{xb}  \br^{(L-1) T}_{x\alpha}=0$,
there are therefore two $x_\beta$-type irreducible representions, namely
\begin{align}
\br^{\Ell T}_{x\beta,1} &= \left( H^\Ell\right)^{-1} \br^{(L-1) T}_{x\alpha,1}\,, \nn\\\
\br^{\Ell T}_{x\beta,2} &= \left( H^\Ell\right)^{-1} \br^{(L-1) T}_{x\alpha,2}
\label{twobetasolns}
\end{align}

where $r^{(L-1)}_{x\alpha,j}$ are given in \eqn{twoxalphasolns}.
For $L=2$ and $L=3$, 
$\br^{\Ell}_{x\beta,1}$ and $\br^{\Ell}_{x\beta,2}$ coincide,
but they are distinct for $L \ge 4$.
\para

\noindent{\bf (3) $x_\gamma$-type null vectors.} 
Having found (for $L \ge 4$)
two irreducible representations of type $x_\alpha$ 
and two of type $x_\beta$,
there must be one remaining,
which we will show to be of the form 
\begin{align}
R^\Ell_{x\gamma} = r^\Ell_{x\gamma} x_\gamma 
\end{align}
where $r^\Ell_{x\gamma}$ is a polynomial in $K$ of maximal degree $L$
(but see below).
Orthogonality to  $\CC^\Ell_{ua}$ and $\CC^\Ell_{ub}$ is automatic.
Orthogonality to $x_a$-type color factors,
$\langle \CC^\Ell_{xa} | R^\Ell_{x\gamma} \rangle  =0$,
implies
\begin{align}
0 &=  \bc^\Ell_{xa} M_{a\gamma} \br^{\Ell T}_{x\gamma}, \qquad 
M_{a\gamma} = \begin{pmatrix}
1 &  0 &  0 & \cdots \\
0 &  1 &  0 & \cdots \\
0 &  0 &  1 & \cdots \\
\vdots & \vdots & \vdots & \ddots \\
\end{pmatrix} 
\label{sonagammafirst}
\end{align}

where $M_{a\gamma}$ is defined in \eqn{innerCR} using the 
${\cal P } $ and ${\cal  Q }$
matrices appropriate to $x_a$ and $x_\gamma$.
One of the $L$-loop $x_a$-type color factors is $c^\Ell_{xa} = K^L$,
so \eqn{sonagammafirst} implies that 
$r^\Ell_{x\gamma}$ is actually of maximal degree $L-1$.
All the other $L$-loop $x_a$-type color factors are obtained from
$(L-1)$-loop $x_b$-type color factors, so that \eqn{sonagammafirst} becomes
\begin{align}
0&=  \bc^{(L-1)}_{xb} G_{ba} \br^{\Ell T}_{x\gamma}
\label{sonagammasecond}
\end{align}

where $G_{ba}$ was defined in \eqn{defGba}.
In addition, orthogonality to $x_b$-type color factors, 
$\langle \CC^\Ell_{xb} | R^\Ell_{x\gamma} \rangle  =0$,
requires
\begin{align}
0& =  \bc^\Ell_{xb} M_{b\gamma} \br^{\Ell T}_{x\gamma}, \qquad 
M_{b\gamma} = \begin{pmatrix}
3 & 1 & 0 &  \cdots \\
0 & 3 & 1 &  \cdots \\
0 & 0 & 3 &  \cdots \\
\vdots & \vdots & \vdots & \ddots \\
\end{pmatrix} 
\label{sonbgamma}
\end{align}
where $M_{b\gamma}$ is defined in \eqn{innerCR} using the 
${\cal P } $ and ${\cal  Q }$
matrices appropriate to $x_b$ and $x_\gamma$.
At one loop, 
\eqns{sonagammasecond}{sonbgamma} are empty,
as there are no $x_b$-type color factors below two loops,
so there is a single $x_\gamma$-type irreducible representation:
\begin{align}
R^\One_{x\gamma} =  x_\gamma \,.
\label{gammaone}
\end{align}

At two loops one has $c^\One_{xb} = K-2$,
so that \eqn{sonbgamma} again yields
a single $x_\gamma$-type irreducible representation:
\begin{align}
R^\Two_{x\gamma} =  (K + \fr{1}{6}) x_\gamma \,.
\label{gammatwo}
\end{align}

For $L \ge 3$, 
one must impose both \eqns{sonagammasecond}{sonbgamma}.
We show in appendix
\ref{sec:appB}
that there is also a single $x_\gamma$-type irreducible representation
that satisfies both \eqns{sonagammasecond}{sonbgamma},
which has the form
\begin{align}
r^\Ell_{x\gamma} =  
\frac{2\left[ 1 - (3K)^L \right] }{3^{L} (1-3K)} 
-
\frac{2\left[ 1 - (-6K)^L \right] }{(-6)^{L} (1+6K)} 
\label{onegammasoln}
\end{align}

consistent with \eqns{gammaone}{gammatwo}.
Again, there is a pattern whereby
$r^\Ell_{x\gamma}$ is given by $K r^{(L-1)}_{x\gamma}$
plus a constant easily obtained from \eqn{onegammasoln}.
\para

\subsection{Summary of null vectors}

\begin{table}[t]
\begin{center}
\begin{tabular}{|l|*{5}{r}|c|}
\hline
\# of loops $L$                   & 0 & 1 & 2  & 3 & 4 &  $L\ge 4$ \\
\hline
\# of $x_\alpha$-type irreps & 0 & 1 & 1 & 2 & 2 &  2 \\
\# of $x_\beta$-type  irreps & 0 & 0 & 1 & 1 & 2 &  2 \\
\# of $x_\gamma$-type irreps & 0 & 1 & 1 & 1 & 1 &  1 \\
\hline
total \# of $x$-type  irreps & 0 & 2 & 3 & 4 & 5 &  5  \\
\hline
\end{tabular}
\end{center}
\caption{
Number of independent $x$-type null vectors for \son.} 
\label{table:sonxnullspace}
\vskip .4 cm
\end{table}

We have explicitly constructed all the $x$-type null
vectors for \son. 
For $L \ge 4$, 
there are ten such null vectors, 
consisting  of five irreducible representations 
whose general forms
are given in eqs.~(\ref{twoxalphasolns}), 
(\ref{twobetasolns}), and (\ref{onegammasoln}).
For $L < 4$, the number of null vectors is fewer
(see table \ref{table:sonxnullspace}).
For the reader's convenience, we explicitly list the
$x$-type null vectors through four loops here:
\begin{align}
\hbox{One loop:   }  
R^\One_{x\alpha} &=  x_\alpha \,, \nn\\
R^\One_{x\gamma} &=  x_\gamma \,, \nn\\[2mm]
\hbox{Two loops:   }  
R^\Two_{x\alpha} &=  (K+ \half) x_\alpha \,, \nn\\
R^\Two_{x\beta} &=   x_\beta  \,, \nn\\
R^\Two_{x\gamma} &=  (K + \fr{1}{6}) x_\gamma \,, \nn\\[2mm]
\hbox{Three loops:   }  
R^\Three_{x\alpha,1} &=  (K^2+ \half K + \fr{1}{4} ) x_\alpha \,, \nn\\
R^\Three_{x\alpha,2} &=  (K^2+ \half K + \fr{5}{36} ) x_\alpha \,, \nn \\
R^\Three_{x\beta} &=  (K + \fr{1}{10}) x_\beta\,, \nn\\
R^\Three_{x\gamma} &= (K^2 + \fr{1}{6}K+ \fr{1}{12}) x_\gamma \,, \nn\\[2mm]
\hbox{Four loops:   }  
R^\Four_{x\alpha,1} &=  (K^3+ \half K^2 + \fr{1}{4}K + \fr{1}{8} ) x_\alpha \,, \nn\\
R^\Four_{x\alpha,2} &=  (K^3+ \half K^2 + \fr{5}{36}K + \fr{11}{216} ) x_\alpha  \,, \nn\\
R^\Four_{x\beta,1} &=  (K^2 + \fr{9}{46} K + \fr{11}{92}) x_\beta \,,\nn\\
R^\Four_{x\beta,2} &=  (K^2 + \fr{57}{382} K + \fr{47}{764}) x_\beta \,,\nn\\
R^\Four_{x\gamma} &= 
(K^2 + \fr{1}{6}K^2+ \fr{1}{12}K + \fr{5}{216}) x_\gamma 
\end{align}
where we have rescaled the $x_\beta$-type null vectors.
\para

\section{$L$-loop \spn\ color space}
\label{sec:spncolor}
\setcounter{equation}{0}

The $L$-loop color space for four-point amplitudes
with gauge group \spn\ can be dealt with summarily 
since the results are nearly identical to those 
for amplitudes with gauge group \son, up to certain relative signs.
The iterative matrices for \spn\ obtained
in sec.~\ref{sec:iterative} are given by
\begin{align}
g_1 &= \begin{pmatrix} K & 0 \cr 0 & K \end{pmatrix}, \quad
g_{xx} = \begin{pmatrix} K & 0 \cr -4 & 0 \end{pmatrix}, \quad
g_{xu} = \begin{pmatrix} K+6 & 3 \cr 0 & -2K \end{pmatrix}, \quad
g_{ux} = \begin{pmatrix} K+3 & -3 \cr 12 & -2K \end{pmatrix}
\label{gspn}
\end{align}

where we have chosen to set $e=2$ and have
expressed these matrices in terms of the \spn\ quadratic Casimir 
$K = N+2 $.
These matrices generate a basis
consisting of polynomials multiplied by
one of four specific (linearly independent) types: 
\begin{align}
x_a      & \equiv [1,0]\otimes \xxi,  \nn \\[1mm]
x_b      & \equiv [K-3,1]\otimes \xxi, \nn \\[1mm]
u_a      & \equiv [K+6,3]\otimes u,  \nn \\[1mm]
u_b      & \equiv [(K-3)(K+6),K-9]\otimes u \,.
\label{spntypes}
\end{align}
Carrying out manipulations exactly analogous to those for \son\ in sec.~\ref{sec:soncolor}, we determine a complete set of color factors that span the 
$L$-loop color space for \spn:
\begin{align}
\CC^\Ell_{xa} &= c^\Ell_{xa} x_a,
& c^\Ell_{xa}& 
\in  \{ K^L \} \cup \{ (K+1)(K-4) c^{(L-1)}_{xb} \} \,, \nn\\
\CC^\Ell_{xb} &= c^\Ell_{xb} x_b,
& c^\Ell_{xb}&
\in \{ K^{n_1} (K+6)^{n_2} (K+2)^{n_3+1} \ | \ n_1 + 2n_2 + 3 n_3 = L-2 \} \,, \nn\\
\CC^\Ell_{ua} &= c^\Ell_{ua} u_a,
& c^\Ell_{ua}& \in \{ c^{(L-1)}_{xa}\}  \,, \nn\\
\CC^\Ell_{ub} &= c^\Ell_{ub} u_b,
& c^\Ell_{ub}& \in \{ c^{(L-1)}_{xb} \} 
\label{spncolor}
\end{align}

which is the same as \eqn{soncolor} up to certain relative signs.
\para

The orthogonal complement of the space of color factors (\ref{spncolor})
is the $L$-loop null space, which (for $L \ge 5$) is spanned by 
ten $x$-type null vectors and seven $u$-type null vectors.
Again carrying out manipulations exactly analogous to those for \son\ 
in sec.~\ref{sec:sonnull}, 
we may determine the explicit forms of all the $x$-type null vectors,
which may be obtained from the \son\ $x$-type null vectors
(\ref{twoxalphasolns}), (\ref{twobetasolns}), and (\ref{onegammasoln}) 
by some obvious changes of relative signs.
\para

\section{Conclusions}
\label{sec:concl}

In this paper, we have analyzed the spaces of color factors 
associated with $L$-loop four-point amplitudes
of fields transforming in the adjoint representation of gauge groups
\sun, \son, or \spn\ by decomposing them into an extended trace basis.
The extended trace basis consists of traces 
(and products of traces) of generators
multiplied by various powers of $N$ 
(or of $K$, where $K$ is proportional
to the quadratic Casimir, 
\viz, $N$ for \sun, $N-2$ for \son, and $N+2$ for \spn).
The dimension of the $L$-loop extended trace space is 
$3L+3$ for \sun\ and $6L+3$  for \son\ and \spn,
and the $L$-loop color space spans a proper subspace 
of the $L$-loop trace space.
Using a refined iterative process, 
we have determined the dimensions of this subspace
for all values of $L$ for the groups \sun, \son, or \spn,
with the results listed in tables 
\ref{table:sunnull} and \ref{table:sonnull}.
We observe that the dimensions of these color spaces
are the same for all these groups up through four loops,
but begin to differ for $L \ge 5$.
\para

As can be seen in
tables \ref{table:sunnull} and \ref{table:sonnull},
the codimensions of the color spaces
(vis-a-vis the extended trace space) reach a fixed value 
for sufficiently large $L$. 
Thus these spaces are more efficiently characterized by specifying
the null space, \ie, the orthogonal complement of the 
color space in the trace space.
Moreover, the null vectors are directly related to group-theory
constraints on the color-ordered amplitudes, as described in
the introduction.
We established the number of null vectors to be four 
for \sun\ (for $L \ge 2$)
and seventeen for \son\ and \spn\  (for $L \ge 5$).
For \sun, we confirmed the forms of the four null vectors 
(or constraints) found previously.
For \son\  (and \spn), we derived explicit expressions for ten
of the seventeen null vectors, namely, the $x$-type null
vectors.  
Obtaining the remaining seven $u$-type null vectors is left for
future work.
\para

Admittedly the usefulness of the null vectors for \son/\spn\ is limited
because they are constructed with respect to an unconventional inner
product.  One might ask why we bother to construct these null vectors
explicitly.   The answer is that proving the existence of these null
vectors, which we do by constructing them, is crucial to establish
the completeness of the basis of $6L-14$ color factors (for $L \ge 5$)
for \son\  constructed in sec.~\ref{sec:soncolor} and listed in table
\ref{table:soncolorspace}.  As explained in sec.~\ref{sec:soncolor}, our
iterative procedure produces the correct number ($L-2$) of independent
$x_b$-type irreducible representations through seven loops, but
(apparently) produces additional ones for $L \ge 8$, corresponding to
solutions of \eqn{integercondition} with $n_3 \ge 2$.  To show that these
additional irreps are not independent of the others, we demonstrate in
appendix \ref{sec:appB} that they are orthogonal to two $x_\alpha$-type
irreps, which we explicitly construct.  There may be other ways to
demonstrate the completeness of the color factors, but this is how
we have done it.  As usual, this is subject to the assumption, stated
in the introduction, that the $L$-loop color space can be obtained by
attaching rungs between any two external legs of the set of $(L-1)$-loop
color factors; it would be nice to have a proof of this assumption.
\para

Another obvious target for future work is 
the characterization of the color spaces
and the null vectors for five-point (and higher) 
amplitudes of \son\ and \spn. 
These were previously found for \sun\ in 
refs.~\cite{Edison:2011ta,Edison:2012fn}. 
\para

\section*{Acknowledgments}
SN would like to thank Jake Bourjaily for helpful conversations.
AO acknowledges a Burns Fellowship from Bowdoin College for
summer support.
This material is based upon work supported by the
National Science Foundation
under Grant No.~PHY21-11943.

\vfil\break

\appendix

\section{Group theory identities}
\label{sec:appA}
\setcounter{equation}{0}

Let $T^a$ denote generators in the defining representation
of \sun, \son, or \spn,
a set of $N \times N$ traceless hermitian matrices 
that in the case of \son\ and \spn\ 
satisfy additional conditions (see below).
For \spn, $N$ is even.
\para

The generators are chosen to be orthonormal
\be
\Tr(T^a T^b) = \index \delta^{ab}
\label{index}
\ee
where $\index$ denotes the index of the defining representation.
These matrices obey commutation relations
\be
[T^a, T^b] = \tf^{abc} T^c
\label{commrlns}
\ee

so that \eqns{index}{commrlns} imply\footnote{For 
the groups SU(2) and Sp(2), 
one has $\tf^{abc} = i \sqrt{2 \index} ~\eps^{abc}$,
while for SO(3), $\tf^{abc} = i \sqrt{\index/2} ~\eps^{abc}$.}
\be 
\tf^{abc} =(1/ \index) \Tr( T^a, [T^b,  T^c] )
\ee
which is manifestly totally antisymmetric.
In the main body of the paper,
we adopt the convention  $c=1$
for the index of the defining representation 
which is commonly used in the amplitudes community.
It is not difficult, however, to adapt our results to other conventions
because all of the quantities considered scale homogeneously with $c$.
\para

We now discuss each classical group separately.

\subsection{\sun}

Generators in the defining representation of \sun\ obey \cite{Cvitanovic:1976am} 
\begin{align}
(T^a)_{ij}  (T^a)_{kl}  &=
\index \left(  \delta_{il} \delta_{jk} 
- {1 \over N} \delta_{ij}  \delta_{kl}  \right) \,.
\label{sunfundrln}
\end{align}

Thus for arbitrary products of generators $\AA$ and $\BB$, we have 
\begin{align}
\Tr(\AA T^a) \Tr(\BB T^a)
& = \index \left[ \Tr (\AA\BB) - \oneN \Tr(\AA) \Tr(\BB) \right] \,,
\nn\\
\Tr(\AA T^a \BB T^a)
&= \index \left[ \Tr (\AA) \Tr(\BB) - \oneN \Tr(\AA \BB) \right] \,.
\label{sunrln}
\end{align}
\para

\subsection{\son}

The generators for \son\  satisfy
\begin{align}
(T^a)^T &= -T^a
\label{son}
\end{align}
where $T$ denotes transpose.
That is, they are antisymmetric as well as 
hermitian (and therefore purely imaginary).
\Eqn{son} implies that $T^a$ is traceless.
\para

Generators in the 
defining representation of $\son$ obey \cite{Cvitanovic:1976am} 
\begin{align}
(T^a)_{ij} (T^a)_{kl} &=
{\index \over 2} 
\left(  \delta_{il}  \delta_{jk}  -  \delta_{ik} \delta_{jl} \right) \,.
\label{sonfundrln}
\end{align}

Hence for arbitrary products of generators $\AA$ and $\BB$, we have 
\begin{align}
\Tr(\AA T^a) \Tr(\BB T^a)
& = {\index \over 2} \bigg[ \Tr (\AA\BB) - \Tr(\AA \BB^T)\bigg] \,,
\nn\\
\Tr(\AA T^a \BB T^a)
&= {\index \over 2} \bigg[ \Tr (\AA) \Tr(\BB) - \Tr(\AA \BB^T) \bigg] \,.
\label{firstsonrln}
\end{align}

Using \eqn{son}, we have
\begin{align}
\BB^T = (-1)^{n_\BB} \BB^R
\label{sonR}
\end{align}
where $\BB^R$ denotes the product of generators $\BB$ in reverse order,
and $n_\BB$ denotes the number of factors in $\BB$.
Thus we can recast \eqn{firstsonrln} as \cite{Huang:2016iqf}
\begin{align}
\Tr(\AA T^a) \Tr(\BB T^a)
& = {\index \over 2} \bigg[ \Tr (\AA\BB) - (-1)^{n_\BB} \Tr(\AA \BB^R)\bigg] \,,
\nn\\
\Tr(\AA T^a \BB T^a)
&= {\index \over 2} \bigg[ \Tr (\AA) \Tr(\BB) - (-1)^{n_\BB} \Tr(\AA \BB^R) \bigg] \,.
\label{sonrln}
\end{align}
\para

\subsection{\spn}

The generators for \spn\ satisfy
\begin{align}
(T^a)^T &=  J T^a J 
\label{spn}
\end{align}
where $J$ is an $N \times N$ matrix 
satisfying $J^2 = -1$ and $J^T = -J$,
where $N$ is even.
\Eqn{spn} implies that $T^a$ is traceless.
\para

Generators in the defining representation of \spn\ 
obey \cite{Cvitanovic:1976am} 
\begin{align}
(T^a)_{ij} (T^a)_{kl} &=
{\index \over 2} 
\left(  \delta_{il}  \delta_{jk}  -  J_{ik} J_{jl} \right) \,.
\end{align}

Hence for arbitrary products of generators $\AA$ and $\BB$, we have 
\begin{align}
\Tr(\AA T^a) \Tr(\BB T^a)
& = {\index \over 2} \bigg[ \Tr (\AA\BB) + \Tr(\AA J \BB^T J )\bigg] \,,
\nn\\
\Tr(\AA T^a \BB T^a)
&= {\index \over 2} \bigg[ \Tr (\AA) \Tr(\BB) - \Tr(\AA J \BB^T J) \bigg] \,.
\label{firstspnrln}
\end{align}

Using \eqn{spn}, we have
\begin{align}
\BB^T = (-1)^{n_\BB-1} J \BB^R J
\label{spnR}
\end{align}
and so we can recast \eqn{firstspnrln} as \cite{Huang:2016iqf}
\begin{align}
\Tr(\AA T^a) \Tr(\BB T^a)
& = {\index \over 2} \bigg[ \Tr (\AA\BB) - (-1)^{n_\BB} \Tr(\AA \BB^R)\bigg] \,,
\nn\\
\Tr(\AA T^a \BB T^a)
&= {\index \over 2} \bigg[ \Tr (\AA) \Tr(\BB) + (-1)^{n_\BB} \Tr(\AA \BB^R) \bigg] \,.
\label{spnrln}
\end{align}

\section{Derivation of \son\ null vectors} 
\label{sec:appB}
\setcounter{equation}{0}

In this appendix, we prove the existence 
of two $x_\alpha$-type \son\ irreducible representations of $S_4$,
which establishes the claim made in sec.~\ref{sec:sonnull}
that the $x_b$-type color space is spanned 
by $L-2$ irreducible representations.
We find the explicit form for these null vectors
and also for the single $x_\gamma$-type irreducible representation.
\para

\subsection{$x_\alpha$-type null vectors}

We recall that the null vector
$R^\Ell_{x\alpha} = r^\Ell_{x\alpha} x_\alpha $
must satisfy the condition (\ref{sonbalpha}), 
which is 
\begin{align}
\bc^\Ell_{xb} \br^{\Ell T}_{x\alpha} = 0
\label{nullcondition}
\end{align}
where 
$\CC^\Ell_{xb} = c^\Ell_{xb} x_b$
is an arbitrary $L$-loop $x_b$-type color factor for \son.
We develop a recursive proof to construct
the null vectors.
Recall from \eqn{threeg} that any $L$-loop $x_b$-type
color factor may be expressed in terms of 
lower loop $x_b$-type color factors
\begin{align}
c^\Ell_{xb} = K c^{(L-1)}_{xb}\,, \qquad 
c^\Ell_{xb} = (K-6)  c^{(L-2)}_{xb} \,, \qquad
c^\Ell_{xb} = (K-2)  c^{(L-3)}_{xb} 
\end{align}
via the operators $g_1$, $g_{bb}^\Two$, and $g_{bb}^\Three$.  
It is useful to express these equations in matrix form:
\begin{align}
\bc^\Ell_{xb} &= \bc^{(L-1)}_{xb} G_1 \,, &
G_1 &= \begin{pmatrix} 
	0 & 1 & 0 & 0 & \cdots \\ 
	0 & 0 & 1 & 0 & \cdots \\ 
	0 & 0 & 0 & 1 & \cdots \\ 
\vdots & \vdots & \vdots & \vdots & \ddots\\
	\end{pmatrix} \,, \nn\\
\bc^\Ell_{xb} &= \bc^{(L-2)}_{xb} G_{bb}^\Two  \,, &
G_{bb}^\Two & = \begin{pmatrix} 
	-6 & 1 & 0 & 0 & \cdots \\ 
	0 &-6 & 1 & 0 & \cdots \\ 
	0 & 0 &-6 & 1 & \cdots \\ 
\vdots & \vdots & \vdots & \vdots & \ddots\\
	\end{pmatrix} \,, \nn \\
\bc^\Ell_{xb} &= \bc^{(L-3)}_{xb} G_{bb}^\Three  \,, &
G_{bb}^\Three & = \begin{pmatrix} 
	-2 & 1 & 0 & 0 & \cdots \\ 
	0 &-2 & 1 & 0 & \cdots \\ 
	0 & 0 &-2 & 1 & \cdots \\ 
\vdots & \vdots & \vdots & \vdots & \ddots\\
	\end{pmatrix} \,. 
\label{matrixG}
\end{align}

We may freely write
\begin{align}
\bc^\Ell_{xb} = \bc^\Ell_{xb} \Pi_{L-1}
\label{xbproj}
\end{align}
where $\Pi_L$ is defined in \eqn{defPi},
since $c^\Ell_{xb}$ is an $(L-1)$th degree polynomial.
Using  \eqns{matrixG}{xbproj},
we see that \eqn{nullcondition}
may be replaced by the following three equations 
\begin{align}
\bc^{(L-1)}_{xb} \Pi_{L-2} G_1 \br^{\Ell T}_{x\alpha} &= 0 \,, \nn\\
\bc^{(L-2)}_{xb} \Pi_{L-3} G_{bb}^\Two \br^{\Ell T}_{x\alpha} &= 0 \,, \nn\\
\bc^{(L-3)}_{xb} \Pi_{L-4} G_{bb}^\Three \br^{\Ell T}_{x\alpha} &= 0 \,.
\label{threenullconditions}
\end{align}

We will now construct two independent solutions of \eqn{threenullconditions}.
First define an infinite vector depending on an arbitrary real number $n$:
\begin{align}
\blam_n &= (1, n, n^2, \cdots ) 
\qquad  \implies  \qquad 
\lambda_n = \blam_n \bK^T = {1 \over 1-nK}  \,.
\end{align}

It is easy to check that 
\begin{align}
G_1 \blam_n^T = n \blam_n^T \,, \qquad
G_{bb}^\Two  \blam_n^T = (n-6) \blam_n^T \,, \qquad
G_{bb}^\Three  \blam_n^T = (n-2) \blam_n^T  \,.
\label{GGG}
\end{align}

Next we consider a truncated version
\begin{align}
\blam^\Ell_n &
= {1 \over n^{L-1} } \blam_n \Pi_{L-1} 
= {1 \over n^{L-1} }(1, n, n^2, \cdots, n^{L-1}, 0, 0, \cdots)
\implies 
\blam^{\Ell T}_n = {1 \over n^{L-1} } \Pi_{L-1}  \blam_n^T \,.
\label{deflamElln}
\end{align}

The following relations are easy to check:
\begin{align}
\Pi_{L-2} G_1  \Pi_{L-1} = \Pi_{L-2} G_1 \,, \qquad
\Pi_{L-3} G_{bb}^\Two  \Pi_{L-1} = \Pi_{L-3} G_{bb}^\Two  \,, \qquad
\Pi_{L-4} G_{bb}^\Three  \Pi_{L-1} = \Pi_{L-4} G_{bb}^\Three    \,.
\label{PiGPi}
\end{align}

Using \eqns{GGG}{PiGPi} we observe that
\begin{align}
\Pi_{L-2} G_1  \blam^{\Ell T}_n 
& = \blam_n^{(L-1)T} \,,\nn\\
\Pi_{L-3} G_{bb}^\Two  \blam^{\Ell T}_n 
&= \left( {n-6\over n^2} \right) \blam_n^{(L-2)T} \,,\nn\\
\Pi_{L-4} G_{bb}^\Three  \blam^{\Ell T}_n 
&= \left( {n-2\over n^3} \right) \blam_n^{(L-3)T}  \,.
\label{threerecursive}
\end{align}

We now recursively prove that $\blam^\Ell_n$ satisfies the 
conditions (\ref{threenullconditions}) to be
the $L$-loop null vector $\br^\Ell_{x\alpha}$.
Assuming that $\blam^\Ell_n$ satisfies
the conditions (\ref{nullcondition}) through $L-1$ loops,
we plug \eqn{threerecursive} into \eqn{threenullconditions}
to see that they satisfy those conditions at $L$ loops.
We also need, however, 
to ensure consistency with the base case.
Recall from \eqn{twoloopnullxalpha} 
that at two loops, the null vector is 
\begin{align}
\br^{\Two}_{x\alpha} =  (\half, 1, 0, \cdots) \,.
\label{basecase}
\end{align}

This is satisfied by $\blam^\Two_n$ only when $n=2$, 
so it appears we only have one solution,
$\br^\Ell_{x\alpha,1} = \blam^\Ell_2$.
However, we may also satisfy \eqn{threenullconditions} with 
a linear combination of two different $\blam^\Ell_n$, 
provided that the constants in parentheses in \eqn{threerecursive}
are degenerate, which is the case for $n=3$ and $n=-6$.
Thus, for any values of $A$ and $B$,
the vector
\begin{align}
\br^\Ell_{x\alpha,2} 
= A \blam^\Ell_3
+ B \blam^\Ell_{-6}
\end{align}

satisfies
\begin{align}
\Pi_{L-2} G_1  
\br^{\Ell T}_{x\alpha,2} & = 
\br^{(L-1)T}_{x\alpha,2}
\,,\nn\\
\Pi_{L-3} G_{bb}^\Two  
\br^{\Ell T}_{x\alpha,2}
&= - \fr{1}{3}
\br^{(L-2)T}_{x\alpha,2} 
\,,\nn\\
\Pi_{L-4} G_{bb}^\Three  
\br^{\Ell T}_{x\alpha,2}
&=  \fr{1}{27}
\br^{(L-3)T}_{x\alpha,2}  \,.
\end{align}

Consistency with the base case (\ref{basecase}) requires
$A=\fr{4}{3}$ and $B=-\fr{1}{3}$ so that finally we have 
two solutions of \eqn{threenullconditions}
\begin{align}
\br^\Ell_{x\alpha,1}  &= \blam^\Ell_2\,, \nn\\
\br^\Ell_{x\alpha,2} &= \fr{4}{3} \blam^\Ell_3  - \fr{1}{3} \blam^\Ell_{-6} \,.
\label{twosolns}
\end{align}

From \eqn{deflamElln}, we have
\begin{align}
\lambda^\Ell_n &= \blam^\Ell_n \bK^T = {1- (nK)^L  \over n^{L-1} (1-nK) } 
\label{lamEllnpoly}
\end{align}

so we can conveniently express \eqn{twosolns} as polynomials 
of degree $L-1$:
\begin{align}
r^\Ell_{x\alpha,1} &=  
\br^\Ell_{x\alpha,1}  \bK^T =
\frac{\left[ 1 - (2K)^L \right] }{2^{L-1} (1-2K)} 
\,,\nn\\
r^\Ell_{x\alpha,2} &=  
\br^\Ell_{x\alpha,2}  \bK^T =
\frac{4\left[ 1 - (3K)^L \right] }{3^{L} (1-3K)} 
+
\frac{2\left[ 1 - (-6K)^L \right] }{(-6)^{L} (1+6K)} 
\,.
\end{align}

\subsection{$x_\gamma$-type null vectors} 

We will now establish that the single $x_\gamma$-type irreducible representation
(for $L\ge 1$)
has the form 
\begin{align}
\br^\Ell_{x\gamma}  &= 
\fr{2}{3} \blam^\Ell_3  + \fr{1}{3} \blam^\Ell_{-6} 
= \left( {2 \over 3^L} \blam_3 - {2 \over (-6)^L} \blam_{-6} \right) \Pi_{L-1}
\label{xgammafirst}
\end{align}

corresponding to a polynomial of degree $L-1$
\begin{align}
r^\Ell_{x\gamma} =  
\br^\Ell_{x\gamma}  \bK^T =
\frac{2\left[ 1 - (3K)^L \right] }{3^{L} (1-3K)} 
-
\frac{2\left[ 1 - (-6K)^L \right] }{(-6)^{L} (1+6K)} 
\,.
\end{align}

We may write \eqn{xgammafirst} as
\begin{align}
\br^\Ell_{x\gamma} 
&= \left( {2 \over 3^L} \blam_3 - {2 \over (-6)^L} \blam_{-6} \right) \Pi_{L}
\label{xgammasecond}
\end{align}

since the $K^L$ term vanishes.
For the matrices
$G_{ba}$ and $M_{b\gamma}$ defined in 
\eqns{defGba}{sonbgamma},
one may ascertain that
\begin{align}
\Pi_{L-2} G_{ba}  \Pi_{L} = \Pi_{L-2} G_{ba} 
\,,\qquad
\Pi_{L-1} M_{b\gamma} \Pi_{L} = \Pi_{L-1} M_{b\gamma} 
\label{PiGMPi}
\end{align}

and also that 
\begin{align}
G_{ba}  \blam_n^T = (n+4)(n-1) \blam_n^T 
\,, \qquad
M_{b\gamma}  \blam_n^T = (n+3) \blam_n^T  \,.
\label{GM}
\end{align}

Using \eqn{xgammasecond} in \eqns{PiGMPi}{GM} 
we have
\begin{align}
\Pi_{L-2} G_{ba}  \br^{\Ell T}_{x\gamma}  &= 
\Pi_{L-2} \left( {28 \over 3^L} \blam^T_3 - {28 \over (-6)^L} \blam^T_{-6} \right) 
= {7\over 3} \br^{(L-1) T}_{x\alpha,2}   \,,
\nn\\
\Pi_{L-1} M_{b\gamma} \br^{\Ell T}_{x\gamma}  &= 
\Pi_{L-1} \left( {12 \over 3^L} \blam^T_3 + {6 \over (-6)^L} \blam^T_{-6} \right) 
= 3 \br^{\Ell T}_{x\alpha,2}   \,.
\label{finalgamma}
\end{align}

The conditions 
(\ref{sonagammasecond}) and (\ref{sonbgamma})
for the $x_\gamma$-type null vector
may be written, using \eqn{xbproj}, as 

\begin{align}
0&=  \bc^{(L-1)}_{xb} \Pi_{L-2} G_{ba} \br^{\Ell T}_{x\gamma},
\qquad 
0 =  \bc^\Ell_{xb} \Pi_{L-1} M_{b\gamma} \br^{\Ell T}_{x\gamma}\,.
\label{gammaconditions}
\end{align} 

Finally, using \eqns{finalgamma}{nullcondition},
we see that \eqn{gammaconditions} is satisfied by \eqn{xgammafirst}.
\para

\vfil\break

\end{document}